%% file: SLR.tex
\documentclass[preprint]{elsarticle} 
\biboptions{sort&compress}

\usepackage[hidelinks]{hyperref}
\usepackage{rotating}
\usepackage{placeins}
\include{custom}
\include{numbers}
\include{glossary}

\newcommand{\change}[1]{\textcolor{black}{#1}} 

\journal{Information and Software Technology}

\begin{document}
\makeatletter
\def\ps@pprintTitle{%
	\let\@oddhead\@empty
	\let\@evenhead\@empty
	\def\@oddfoot{\parbox[t]{1\linewidth}{\footnotesize \textit{Accepted Manuscript. \@journal  \hfill December 04, 2021 \\ ~\url{https://doi.org/10.1016/j.infsof.2021.106800} \\
				License: \href{https://creativecommons.org/licenses/by-nc-nd/4.0/}{CC BY-NC-ND 4.0}	}}}%
	\let\@evenfoot\@oddfoot}
\makeatother
\begin{frontmatter}
\title{A Systematic Literature Review on \\ Counterexample Explanation}

\author[mymainaddress,mysecondaryaddress]{Arut Prakash Kaleeswaran}
\author[mymainaddress]{Arne Nordmann}
\author[mysecondaryaddress]{\\Thomas Vogel}
\author[mysecondaryaddress]{Lars Grunske}

\address[mymainaddress]{Bosch Corporate Sector Research, \emph{Renningen, Germany}\\
	\{arutprakash.kaleeswaran$|$arne.nordmann\}@de.bosch.com\\}
\address[mysecondaryaddress]{Humboldt-Universität zu Berlin, \emph{Berlin, Germany}\\ \{thomas.vogel$|$grunske\}@informatik.hu-berlin.de\vspace{-1em}}

\begin{abstract}

\textbf{Context:}
Safety is of paramount importance for cyber-physical systems in domains such as automotive, robotics, and avionics.
Formal methods such as model checking are one way to ensure the safety of cyber-physical systems. However, adoption of formal methods in industry is hindered by usability issues, particularly the difficulty of understanding model checking results.
\textbf{Objective:}
We want to provide an overview of the state of the art for counterexample explanation by investigating the contexts, techniques, and evaluation of research approaches in this field. This overview shall provide an understanding of current and guide future research.
\textbf{Method:}
To provide this overview, we conducted a systematic literature review. The survey comprises \pubfinal publications that address counterexample explanations for model checking.
\textbf{Results:}
Most primary studies 
provide counterexample explanations graphically or as traces,
minimize counterexamples to reduce complexity, 
localize errors in the models expressed in the input formats of model checkers, 
support linear temporal logic or computation tree logic specifications, 
and
use model checkers of the Smybolic Model Verifier family.
Several studies evaluate their approaches in safety-critical domains with industrial applications.
\textbf{Conclusion:}
We notably see a lack of research on counterexample explanation that 
targets probabilistic and real-time systems, 
leverages the explanations to domain-specific models,
and evaluates approaches in user studies.
We conclude by discussing the adequacy of different types of explanations for users with varying domain and formal methods expertise, showing the need to support laypersons in understanding model checking results to increase adoption of formal methods in industry.
\end{abstract}

\begin{keyword}
formal methods, model checking, counterexample explanation
\end{keyword}
\end{frontmatter}

\input{tex/Introduction.tex}
\input{tex/RelatedSurvey.tex}
\input{tex/Process.tex}
\input{tex/Analysis.tex}

\input{tex/ThreatsToValidity.tex}
\input{tex/Discussion.tex}
\input{tex/conclusion}

\setlength{\bibsep}{0pt plus 0.3ex} 
\bibliography{SLR}
\bibliographystyle{elsarticle-num}



\end{document}

%% file: custom.tex
\usepackage[utf8]{inputenc}
\usepackage[american]{babel}
\usepackage{amsmath}
\usepackage{amssymb}
\usepackage{array}
\usepackage{csquotes}

\usepackage{amsthm}
\usepackage{wrapfig}
\usepackage{lscape}
\usepackage{rotating}
\usepackage{epstopdf}
\usepackage{listings}
\usepackage{multirow}
\usepackage[normalem]{ulem}
\useunder{\uline}{\ul}{}
\usepackage{lscape}
\usepackage{longtable}
\usepackage[pass]{geometry}
\usepackage{caption}
\usepackage{subcaption}
\usepackage[inline]{enumitem}
\usepackage[acronym,nonumberlist]{glossaries}
\glsdisablehyper
\usepackage{caption}
\usepackage{subcaption}

\usepackage{todonotes}
\usepackage{setspace}
\usepackage{url}

\usepackage{xspace}

\usepackage{hyphenat}

\newcommand{\cf}[0]{\textit{cf.}\@\xspace}
\newcommand{\ie}[0]{\textit{i.\,e.},\@\xspace}

\newcommand{\eg}[0]{\textit{e.\,g.},\@\xspace}

\newcommand{\etal}[0]{\textit{et\,al.}\@\xspace}

\newcommand{\sect}[1]{Section\xspace\ref{sec:#1}}
\newcommand{\fig}[1]{\figurename\xspace\ref{fig:#1}}

\newcommand{\concept}[1]{\textit{#1}}

\newcommand{\step}[1]{\texttt{Step\,\textcircled{#1}}}

\newcommand{\rsq}[1]{\textbf{RQ#1:}\xspace}
\newcommand{\rsqtext}[1]{\textbf{RQ#1}\xspace}
\newcommand{\rsqparagraph}[2]{~\vspace{-.75em}\par\noindent \rsq{#1} \textit{#2}}

%% file: numbers.tex
\newcommand{\pubsteptwoout}[0]				{12\xspace}

\newcommand{\pubstepthreeout}[0]			{60\xspace}
\newcommand{\pubstepfour}[0]					{672\xspace}
\newcommand{\pubstepfive}[0]					{522\xspace}
\newcommand{\pubstepfiveout}[0]				{150\xspace}
\newcommand{\pubstepsix}[0]						{172\xspace}
\newcommand{\pubstepsixout}[0]				{350\xspace}
\newcommand{\pubstepsixremaining}[0]	{255\xspace}

\newcommand{\pubstepsevenout}[0]			{152\xspace}

\newcommand{\pubfinal}[0]							{116\xspace}

\newcommand{\pubsteponenew}[0]						{168\xspace}
\newcommand{\pubsteptwonew}[0]						{156\xspace}
\newcommand{\pubstepthreenew}[0]					{96\xspace}
\newcommand{\pubstepsixremainingnew}[0]	{268\xspace}

\newcommand{\represgraphical}[0]			{54\xspace}
\newcommand{\represtectual}[0]				{17\xspace}

\newcommand{\represfta}[0]				{7\xspace}
\newcommand{\represfuml}[0]				{27\xspace}

\newcommand{\inoutdep}[0]							{21\xspace}

\newcommand{\inprogram}[0]						{15\xspace}
\newcommand{\outprogram}[0]						{5\xspace}

\newcommand{\insel}[0]								{4\xspace}
\newcommand{\outsel}[0]								{5\xspace}

\newcommand{\outgraph}[0]							{13\xspace}
\newcommand{\outft}[0]								{7\xspace}
\newcommand{\inmlmc}[0]								{35\xspace}

\newcommand{\specltl}[0]							{31\xspace}
\newcommand{\specctl}[0]							{10\xspace}
\newcommand{\specpctl}[0]							{5\xspace}
\newcommand{\speccsl}[0]							{5\xspace}

\newcommand{\toolsnusmv}[0]						{31\xspace}
\newcommand{\toolsspin}[0]						{12\xspace}
\newcommand{\toolsmaude}[0]						{5\xspace}
\newcommand{\toolsvis}[0]							{4\xspace}
\newcommand{\toolsprism}[0]						{15\xspace}
\newcommand{\toolsmrmc}[0]						{3\xspace}
\newcommand{\fwautofocus}[0]					{3\xspace}
\newcommand{\fwmodchk}[0]							{4\xspace}
\newcommand{\fwassert}[0]							{4\xspace}
\newcommand{\fwkegvis}[0]							{3\xspace}
\newcommand{\toolsreuse}[0]						{4\xspace}

\newcommand{\processminim}[0]					{38\xspace}
\newcommand{\processwitness}[0]				{9\xspace}
\newcommand{\processmultip}[0]				{7\xspace}
\newcommand{\represfsearch}[0]				{14\xspace}

\newcommand{\evaldomaincomm}[0]				{14\xspace}

\newcommand{\evaldomainautomotive}[0]	{10\xspace}
\newcommand{\evaldomainnuclear}[0]		{4\xspace}
\newcommand{\evaldomainavionics}[0]		{5\xspace}
\newcommand{\evaldomainrobotics}[0]		{6\xspace}
\newcommand{\evaldomainrailway}[0]		{3\xspace}

\newcommand{\evalusecaseind}[0]				{16\xspace}
\newcommand{\evalusecaseindref}[0]		{22\xspace}
\newcommand{\evalusecasenonind}[0]		{7\xspace}
\newcommand{\evalusecasenonindref}[0]	{31\xspace}
\newcommand{\evalusecaseexample}[0]		{17\xspace}

\newcommand{\evalaspectperform}[0]		{55\xspace}
\newcommand{\evalaspecteffect}[0]			{51\xspace}
\newcommand{\evalaspectscale}[0]			{13\xspace}
\newcommand{\evalaspectperformeffect}[0]{6\xspace}

\newcommand{\evalmethodbenchm}[0]			{21\xspace}
\newcommand{\evalmethodusecase}[0]		{85\xspace}
\newcommand{\evalmethoduserstudy}[0]	{6\xspace}

\newcommand{\tabdataitems}[0]	{\cite[Table\,1]{appendix}\xspace}
\newcommand{\tabrqone}[0]	{\cite[Table\,2]{appendix}\xspace}
\newcommand{\tabrqtwo}[0]	{\cite[Table\,3]{appendix}\xspace}
\newcommand{\tabrqthree}[0]	{\cite[Table\,6]{appendix}\xspace}

\newcommand{\tabrqfourab}[0]	{\cite[Tables\,7 and 8]{appendix}\xspace}

\newcommand{\tabrqfourc}[0]	{\cite[Table\,9]{appendix}\xspace}

\newcommand{\tabrqfivea}[0]	{\cite[Table\,10]{appendix}\xspace}
\newcommand{\tabrqfiveb}[0]	{\cite[Table\,11]{appendix}\xspace}

\newcommand{\tabrqsixa}[0]	{\cite[Table\,12]{appendix}\xspace}
\newcommand{\tabrqsixb}[0]	{\cite[Table\,13]{appendix}\xspace}

\newcommand{\tabrqsevena}[0]	{\cite[Table\,14]{appendix}\xspace}
\newcommand{\tabrqsevenb}[0]	{\cite[Table\,15]{appendix}\xspace}
\newcommand{\tabrqsevenc}[0]	{\cite[Table\,16]{appendix}\xspace}
\newcommand{\tabrqsevend}[0]	{\cite[Table\,17]{appendix}\xspace}

%% file: glossary.tex
\makenoidxglossaries

\newacronym{BFL}{BFL}{Brute Force Lifting}
\newacronym{GF}{GF}{Grammatical Framework}
\newacronym{MPS}{MPS}{Meta Programming System}
\newacronym{XBF}{XBF}{eXtended Best-First}
\newacronym{DFS}{DFS}{Depth-First Search}
\newacronym{BFS}{BFS}{Breath-First Search}
\newacronym{FMEA}{FMEA}{Failure Mode and Effect Analysis}
\newacronym{FTA}{FTA}{Fault Tree Analysis}
\newacronym{HAZOP}{HAZOP}{Hazard and Operability}
\newacronym{UML}{UML}{Unified Modeling Language}
\newacronym{SysML}{SysML}{Systems Modeling Language}
\newacronym{LTL}{LTL}{Linear Temporal Logic}
\newacronym{CTL}{CTL}{Computation Tree Logic}
\newacronym{SAT}{SAT}{Satisfiability}
\newacronym{BDD}{BDD}{Binary Decision Diagrams}
\newacronym{PSL}{PSL}{Property Specification Language}
\newacronym{RTCTL}{RTCTL}{Real Time Computation Tree Logic}
\newacronym{TCTL}{TCTL}{Timed Computational Tree Logic}
\newacronym{PLC}{PLC}{Programmable Logic Controller}
\newacronym{PCTL}{PCTL}{Probabilistic Computation Tree Logic}
\newacronym{CSL}{CSL}{Continuous Stochastic Logic}
\newacronym{CL}{CL}{Contract Language}
\newacronym{DTMC}{DTMC}{Discrete-Time Markov Chain}
\newacronym{CTMC}{CTMC}{Continuous-Time Markov Chain}
\newacronym{MDP}{MDP}{Markov Decision Processes}
\newacronym{MSC}{MSC}{Message Sequence Chart}
\newacronym{SMT}{SMT}{Satisfiability Modulo Theories}
\newacronym{MILP}{MILP}{Mixed Integer Linear Programming}
\newacronym{GUI}{GUI}{Graphical User Interface}
\newacronym{DSL}{DSL}{Domain-Specific Language}
\newacronym{SMV}{SMV}{Symbolic Model Verifier}
\newacronym{NuSMV}{NuSMV}{New Symbolic Model Verifier}
\newacronym{SPIN}{SPIN}{Simple PROMELA Interpreter}
\newacronym{PROMELA}{PROMELA}{Process or Protocol Meta Language}
\newacronym{VIS}{VIS}{Verification Interacting with Synthesis}
\newacronym{ACL2}{ACL2}{A Computational Logic for Applicative Common Lisp}
\newacronym{ASSERT}{ASSERT}{Analysis of Semantic Specifications and Efficient generation of Requirements-based Tests}
\newacronym{MRMC}{MRMC}{Markov Reward Model Checker}
\newacronym{PRISM}{PRISM}{Probabilistic Symbolic Model Checker}
\newacronym{CLAN}{CLAN}{Contract Language ANalyser}
\newacronym{FASTEN}{FASTEN}{FormAl SpecificaTion ENvironment}
\newacronym{DiPro}{DiPro}{Directed Probabilistic Counterexample Generation Tool}
\newacronym{KEGVis}{KEGVis}{Kounterexample generator and visualizer}
\newacronym{XChek}{XChek}{Multi-valued Model-Checker}
\newacronym{AMASE}{AMASE}{Aerospace Multi-agent Simulation Environment}
\newacronym{UAV}{UAV}{Unmanned Aerial Vehicle}
\newacronym{STANCE}{STANCE}{Structural Analysis of Counter-Examples}
\newacronym{COMICS}{COMICS}{Computing Minimal Counterexamples}
\newacronym{GraphML}{GraphML}{Graph Markup Language}
\newacronym{RAE}{RAE}{Requirements Analysis Engine}
\newacronym{CBMC}{CBMC}{C Bounded Model Checker}
\newacronym{BPMN}{BPMN}{Business Process Model and Notation}
\newacronym{CNL}{CNL}{Controlled/Constrained Natural Language}
\newacronym{CAD}{CAD}{Computer-aided Design}
\newacronym{Butramin}{Butramin}{BUg TRAce MINimization}
\newacronym{CBD}{CBD}{Contract-Based Design}
\newacronym{LTS}{LTS}{Labelled Transition System}


%% file: tex/Introduction.tex
\section{Introduction}
\label{sec:intro}
\vspace{-.5em}
\begin{center}
	\blockquote{\textit{It is impossible to overestimate the importance of the counterexample feature. The counterexamples are invaluable in debugging complex systems. Some people use model checking just for this feature.} -- Edmund Clarke~\cite{Clarke08}}
\end{center}

\noindent
The complexity of modern, software-intensive systems continues to increase due to the rising number of features and functionalities. When complex software-intensive systems are used in safety-critical domains such as automotive, robotics, and avionics, their malfunction might lead to severe damages or even loss of lives. Consequently, safety of these systems is of paramount importance. To ensure safety, these systems have to be developed according to standards such as IEC~61508, and ISO~26262 in the automotive domain. These standards require safety methods such as \gls{FMEA}, \gls{FTA}, or \gls{HAZOP}.
Still today, such safety analysis is often performed manually by engineers to identify potential safety flaws. Machine support to automate the analysis process is a way to overcome this time-consuming, expensive, and error-prone process.

Model checking~\cite{ClarkeGP01,0020348,ClarkeHV18} is a computer-assisted verification method for systems that were modeled in a formal way by state-transition systems. Drawing from research in mathematical logic, programming languages, hardware design, and theoretical computer science, model checking is now widely used for the verification of hardware and software in industry~\cite{ClarkeHV18}.

Model checking verifies whether a requirement is satisfied by a system or not. For this purpose, it requires a formal specification of the requirement, typically as a \concept{temporal-logic formula}~$\varphi$, and a formal description of the system, for instance, as a \concept{Kripke structure}~$K$.
Then, a \concept{model checker} as a tool performs the verification by checking whether the specification $\varphi$ is satisfied by the system model $K$, that is, $K$~$\vDash$~$\varphi$.
The result of the model checking is either that
$\varphi$ is satisfied by $K$ (\ie~$K$~$\vDash$~$\varphi$), or that
$\varphi$ is not satisfied by $K$ (\ie~$K$~$\nvDash$~$\varphi$).
In the latter case, the model checker returns a \concept{counterexample} to $\varphi$ on $K$. Such a counterexample describes an execution path over system states that leads from the initial state to a state that violates  $\varphi$. Each state of such a path consists of atomic propositions ($AP$) over the variables defined by $K$.

Once the system and requirements are formalized, model checking is attractive as it is an automated method and offers counterexamples if a system model fails to satisfy a requirement, serving as indispensable debugging information~\cite{0020348}.
However, counterexamples are only the symptoms of faults and understanding a counterexample to identify a fault in the system model is a complicated task for several reasons:
\begin{enumerate*}[label={(\textbf{R\arabic*})},itemjoin={{, }}, itemjoin*={{, and }}]
	\item a counterexample is often cryptic and lengthy~\cite{MuramTZ15}
	\item not all the states in a counterexample are relevant to an error~\cite{BergSJ07}
	\item not all the variables in a state have any relation to the violated specification~\cite{BergSJ07}
	\item the debugging task is performed manually, which is time-consuming and error-prone~\cite{BarbonLS18,MuramTZ15,Ratiu2021,OvsiannikovaBPV21}
	\item the counterexample does not explicitly highlight the source of the error that is hidden in the model~\cite{BarbonLS18}.
\end{enumerate*}
These challenges call for a method to explain counterexamples, assisting system designers in localizing faults in their models~\cite{LeueB12}.

Therefore, we provide an overview of the state of the art in research on explaining counterexamples and how system engineers are supported in interpreting counterexamples. This allows us to assess the state of the art and identify needs for future work.
For this purpose, we conducted a system literature review on counterexample explanation with the main focus on
\begin{enumerate*}[label={(\textbf{\arabic*})},itemjoin={{, }}, itemjoin*={{, and }}]
	\item the different kinds to explain a counterexample
	\item the different methods to transform or optimize a counterexample in order to provide an explanation
	\item influences of the input system and requirement on counterexample explanation
	\item the different domains and applications to evaluate approaches to counterexample explanation.
\end{enumerate*}

\begin{figure}[b!]
	\centering
	\includegraphics[width=.8\textwidth]{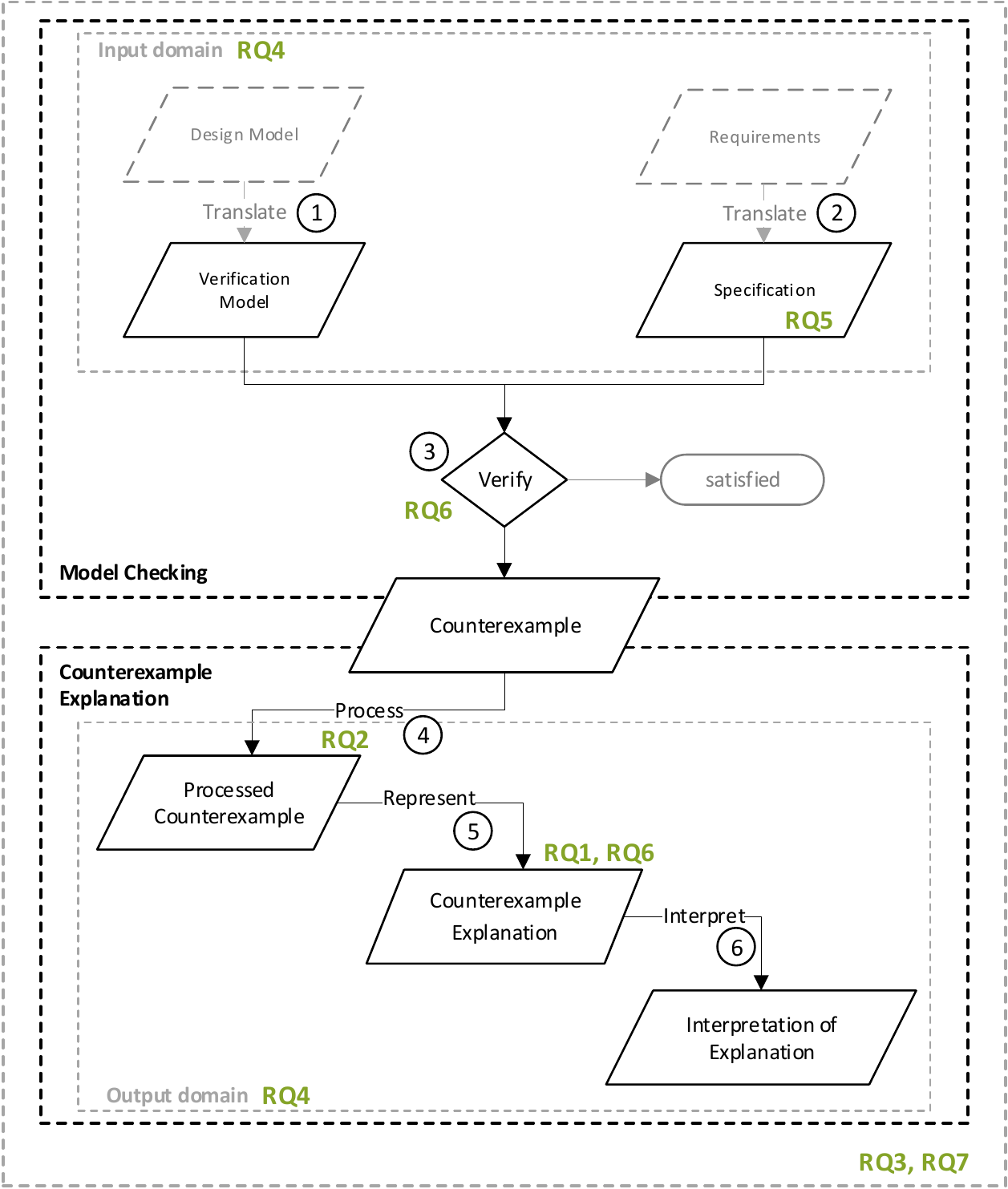}
	\caption{Overview of model checking and counterexample explanation.}
	\label{fig:overview}
\end{figure}

By collecting and analyzing literature for these aspects, our survey provides a comprehensive overview of counterexample explanation. To structure and guide our survey, introduce the conceptual model with its terminology that we use in our survey, depicted in \fig{overview}.

\subsection{Conceptual Model and Terminology}
\label{sec:intro:terminology}

\noindent
The conceptual model shown in \fig{overview} consists of two parts, \textit{model checking} and \textit{counterexample explanation}, that we discuss in the following.

\paragraph{Model Checking}

\noindent
In our survey, we use the term \concept{design model} to refer to an informal or formal description of the system that can be expressed in any modeling language such as the \gls{SysML} and \gls{UML}.
To perform model checking, the design model has to be translated to a \concept{verification model} that is expressed in a specific formalism required by the used model checker (\step{1} in \fig{overview}).
For instance, the model checker \gls{NuSMV}~\cite{CimattiCGR99,CimattiCGGPRST02} uses the formalism of a Kripke structure to describe a system.

As a system specification we generally consider technical and system \concept{requirements} that are usually defined informally. To perform model checking, requirements have to be formally specified by translating them into a \concept{specification} expressed in a temporal logic (\step{2}). For instance, \gls{NuSMV} supports, among others, \gls{LTL}~\cite{Pnueli77} and \gls{CTL}~\cite{ClarkeE81}.
To ease this formalization step, property specification patterns have been proposed~\cite{DwyerAC99,AutiliGLPT15}.

Given a verification model and specification, a model checker \concept{verifies} whether the model satisfies the specification (\step{3}). The result of the verification is either the fact that the model \concept{satisfies} the specification or a \concept{counterexample} showing a sequence of state transitions of the model that violates the specification: the counterexample. Such a counterexample is produced in a specific format depending on the model checker. For instance, a counterexample created by \gls{NuSMV} follows the Kripke structure of the verification model.

\paragraph{Counterexample Explanation}

\noindent
The goal of counterexample explanation is to support engineers in interpreting a counterexample and comprehending the violation and particularly, the error in the system design.

The input to the counterexample explanation process is the \concept{counterexample} returned by a model checker and optionally the artifacts of what we call the \concept{input domain} comprising the \concept{verification model}, \concept{specification}, \concept{design model}, and \concept{requirements} of the system.
First, the \concept{counterexample} is \concept{processed} to yield a \concept{processed counterexample} that may be easier to interpret (\step{4}). For instance, it can be a minimized version of the original counterexample reduced to the parts that are relevant for the violation and error.
Afterwards, a \concept{representation} of the processed counterexample is created that serves as the \concept{explanation} of the counterexample (\step{5}).
The explanation may relate to artifacts of the input domain, which have been provided by the system engineer.
Finally, the explanation is presented to a system engineer who \concept{interprets} it, usually manually, to comprehend the error (\step{6}).

%% file: tex/RelatedSurvey.tex
\section{Related Work}
\label{sec:related_work}

\noindent
In the following, we discuss surveys related to our work, categorized into two groups depending on whether they are focused on verification for specific application domains or model checking and its algorithms.

\paragraph{Surveys focused on verification for specific application domains}
Karna~\etal~\cite{KarnaCYZZ18} present a survey about model checking and various tools or techniques used in software engineering development.
An overview of formal verification on real-time systems is given by Wang~\cite{Wang04} and on model-based engineering by Gabmeyer~\etal~\cite{GabmeyerKSGK19}. Both surveys discuss models, specification languages, verification frameworks, and the state-space construction and representation in their respective domains.
More specifically, Ovatman~\etal~\cite{OvatmanAPU16} discuss the difficulties of the model checking process, tools and specification language on \gls{PLC} software production.
The survey by Grimm~\etal~\cite{Grimm2018} provides an overview of the most widely used formal verification techniques and discusses their value for critical Systems-on-Chip verification.

\paragraph{Surveys focused on model checking techniques and its algorithms}
The surveys by Clarke~\etal~\cite{ClarkeGJLV01,ClarkeV03} present contributions on symbolic model checking with its algorithms, abstraction, and software verification.
The survey by Prasad~\etal~\cite{PrasadBG05} investigates \gls{SAT}-based
model checking, bounded model checking, and problems such as application-specific heuristics and conflict-driven learning.
Another survey focusing on SAT-based model checking but together with hardware benchmarks is given by Amla~\etal~\cite{AmlaDKKM05} for eight bounded and unbounded techniques.
The survey by D'Silva~\etal~\cite{DSilvaKW08} focuses on automatic static analysis of software for three techniques, namely static analysis with abstract domains, model checking, and bounded model checking.
Finally, there are surveys discussing the problem of state space explosion in model checking. For instance, Pelanek~\cite{Pelanek08} survey the techniques to address the state space explosion. They categorize such techniques to state space reductions, storage size reductions, parallel and distributed computation, randomization and heuristics.
The survey by Edelkamp~\etal~\cite{EdelkampSBWFA08} discusses algorithms for directed model checking and mitigating the state explosion problem.

\paragraph{Summary}
Given the existing surveys in these two categories, none of them addresses the research problem of explaining counterexamples generated by a model checker.
Therefore, to the best of our knowledge, this systematic literature review is the first one that investigates specifically the state of the art in explaining counterexamples in order to support debugging and localizing faults.

%% file: tex/Process.tex
\section{Research Methodology}
\label{sec:process}

\noindent
For our survey, we followed guidelines from standard practice in systematic literature reviews by Kitchenham and Charters~\cite{KitchenhamS07} complemented by guidelines on snowball sampling by Wohlin \etal~\cite{WohlinRHOR00}. Particularly, we follow these major steps to conduct a systematic literature review~\cite{KitchenhamS07}:
\begin{enumerate*}[label={(\textbf{\arabic*})},itemjoin={{, }}, itemjoin*={{, and }}]
	\item define the need for the review
	\item define research questions
	\item identify primary studies
	\item perform data extraction
	\item perform data synthesis
	\item study quality assessment. 
\end{enumerate*}
Concerning the first step, we have motivated the need of our survey in the introduction and the necessity of this study based on related work (\sect{related_work}).

\subsection{Research Questions}
\label{sec:RQ}

\noindent
We argue that counterexample explanation is key to a broader adoption of model checking because it allows the usage of model checkers without necessarily having to understand their internal intricacies. Moreover, it promises to make the model checking result actionable for engineers by supporting them to localize and understand the fault in the system design.
The main motive of this survey is therefore to address the general question: \textit{What is the state of the art for explaining counterexamples generated by model checkers?}

To answer the general question, we refined it into eight specific research questions (RQs). \fig{overview} positions these research questions in our conceptual overview of model checking and counterexample explanation. 

\rsqparagraph{1}{How are counterexamples explained and what are the effects of this explanation on counterexample interpretation?}
With this question, we target the explanation of counterexamples presented to engineers, their style of representation, and their impact on the interpretation of counterexamples.

\rsqparagraph{2}{How are counterexamples processed and what effect does it have on interpreting the counterexample?}
Answering this RQ provides insights into the effects of processed counterexamples for their interpretation.

\rsqparagraph{3}{What kind of additional information is used to enrich the counterexample explanation?}
This RQ investigates additional information that is provided together with the counterexample explanation to improve debugging and locating the error.

\rsqparagraph{4}{Which input domains are targeted by counterexample explanation approaches? What is the influence of the input domain on the explanation?}
This RQ explores the relation between the input domain and output domains of counterexample representation, including the influence of the input domain on the output domain.

\rsqparagraph{5}{What are the different temporal logics used to express system specifications and what type of properties are covered in counterexample explanation approaches?} With this RQ, we enumerate and count the temporal logic formalisms being used to specify properties in counterexample explanation approaches. Moreover, we look at the types of properties supported by these approaches, \eg safety and liveness.

\rsqparagraph{6}{Which verification tools and frameworks are developed and used to explain counterexamples, and how do they effect the counterexample explanation?}
This RQ lists different verification tools and frameworks and further describes their impact on representing and explaining counterexamples.

\rsqparagraph{7}{How are counterexample explanation approaches evaluated?}
This RQ provides insights into the evaluation of counterexample explanation approaches, \ie the evaluated aspects, evaluation methods, domains, and applications.

~\\
Considering \fig{overview} that positions all of the research questions in a conceptual overview of model checking and counterexample explanation, we note that the research questions tackle all relevant elements. Therefore, we are confident that our survey provides a comprehensive overview of the state of the art in counterexample explanation.

\subsection{Selection of Primary Studies}
\label{sec:iden_process}

\begin{figure}[t!]
	\centering
	\includegraphics[width=\textwidth]{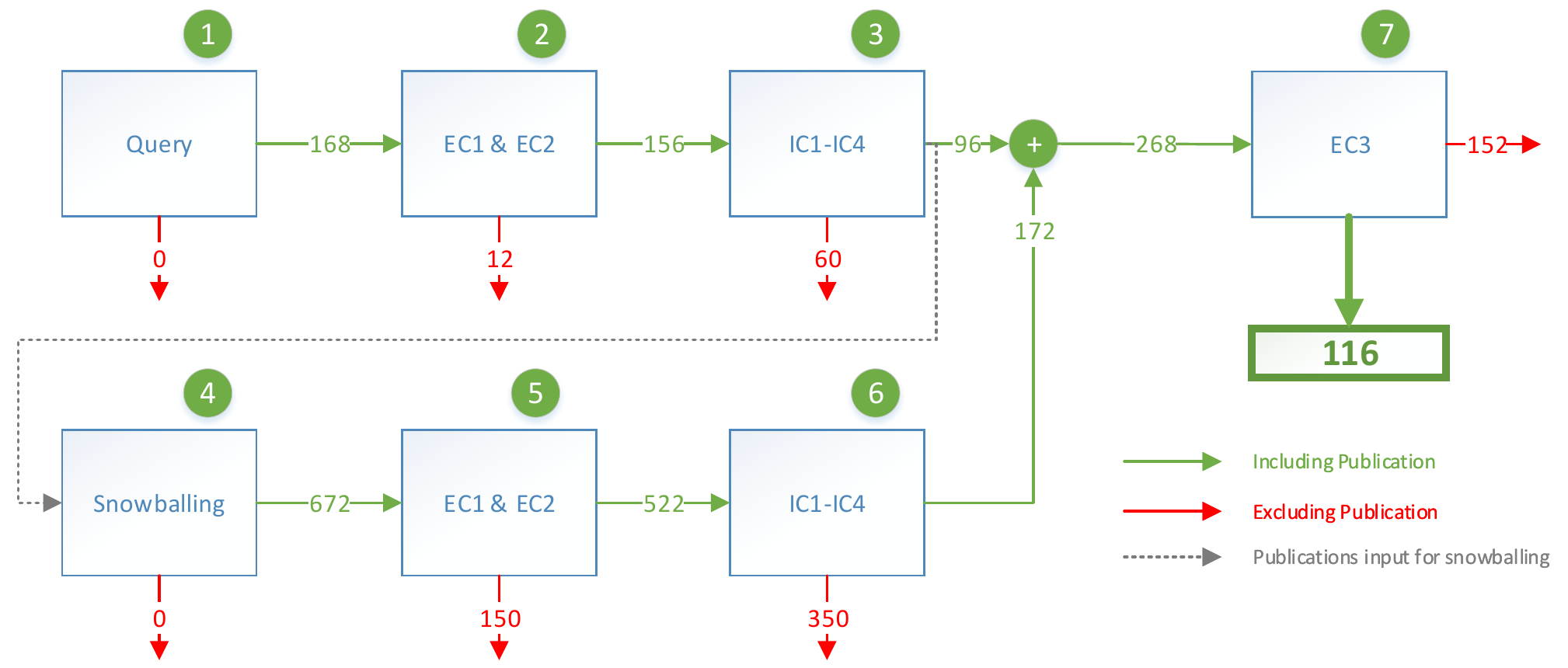}
	\caption{Process to identify the primary studies for our survey.}
	\label{fig:process}
\end{figure}

\noindent
We select primary studies in two phases.
First, we collect the initial set of primary studies by a keyword-based search and filter them by applying the inclusion and exclusion criteria. Second, we perform snowballing with the initial set and again filter the newly identified primary studies using defined inclusion and exclusion criteria.
The selection process of primary studies is shown in \fig{process}.

\subsubsection{Keyword-Based Search}
\label{sec:startset}

\change{Using keyword-based search, we identify a start set for the subsequent snowballing. We use \emph{Google Scholar} with the search query shown in \autoref{query} and limit the results to publications published between 2000 and 2021. We performed the query on February 11th, 2021.}
\begin{lstlisting}[caption={Query for the keyword-based search.},label=query,basicstyle=\ttfamily,columns=fullflexible,frame=single,breaklines=true]
(Abstraction|Annotated|Graphical|Localization|Explanation|Guided|Interpretation|Lifting|Minimum|Refinement|Pattern|Highlight|Reduced|safety|Liveness|Reachability|CNL|DSL|Causality|Ontology)(Counterexample|Model Checker|Temporal Logic)
\end{lstlisting}

We derive the keywords of the query from the survey's focus: explanation and interpretation of counterexamples. Then \emph{2dSearch}\footnote{\url{https://app.2dsearch.com/query}} generates the combination of the keywords to a query that allows to combine the keywords visually and check whether Google Scholar supports specific combinations.
We further use the tool \emph{Publish or Perish}\footnote{\url{https://harzing.com/resources/publish-or-perish}} that performs the search and automatically extracts the search results from Google Scholar.
\change{The query yields \textbf{\pubsteponenew} primary studies, see \step{1} in \fig{process}.
}

\subsubsection{Applying Inclusion and Exclusion Criteria}
\label{sec:icec}

\noindent
All publications resulting from the query are analyzed manually based on the following inclusion and exclusion criteria. This analysis is conducted with a four-eyes principle where one author assesses the analysis of another author.

A primary study is included in our study if it satisfies at least one of the following inclusion criteria (\textbf{IC}).  
\begin{enumerate}[label={(\textbf{IC\arabic*})}]\itemsep0em
	\item The primary study describes an approach
	to represent counterexamples either in a graphical or prose manner. 
	\item The primary study describes an approach
	that
	modifies the length of a counterexample to ease counterexample explanation or interpretation.
	\item The primary study describes a framework to represent counterexamples.
	\item The primary study describes the problem statements or challenges of explaining or interpreting counterexamples.
\end{enumerate}

\noindent
Further, we exclude any primary study if it satisfies at least one of the following exclusion criteria \textbf{(EC)}.
\begin{enumerate}[label={(\textbf{EC\arabic*})}]\itemsep0em
	\item The primary study is a book, a PhD thesis, or a patent.
	\item The primary study is not accessible online from the Bosch or Humboldt University network.
\end{enumerate}

\noindent
In general, we first apply the exclusion and then the inclusion criteria on the publication's title, abstract, introduction, and conclusion.

\change{Applying the two EC on the \textbf{\pubsteponenew} primary studies obtained by the query (\step{2} in \fig{process}) filters out \textbf{\pubsteptwoout} studies, leaving \textbf{\pubsteptwonew} remaining studies.
Based on the IC, we exclude \textbf{\pubstepthreeout} primary studies and include \textbf{\pubstepthreenew} studies (\step{3}). These \textbf{\pubstepthreenew} studies are the input for snowballing.}

\subsubsection{Snowballing}
\label{sec:snowballing}

\noindent
\change{In the second phase, we perform snowballing with the primary studies that we identified in the first phase. Particularly, we use Google Scholar to perform forward (identifying citing publications) and a backward (identifying cited publications based on the publication's references and related articles suggested by Google Scholar) snowball sampling. Then, exclusion and inclusion criteria are applied to the newly discovered primary studies as well. Snowballing is performed by the first author of this paper and the other authors verify it by randomly selecting primary studies and verifying the snowball sampling.}

\change{We collect additional \textbf{\pubstepfour} primary studies during snowballing with the \textbf{\pubstepthreenew} primary studies of the first phase (\step{4} in \fig{process}).
Applying the two EC on these \textbf{\pubstepfour} studies (\step{5}) excludes \textbf{\pubstepfiveout} studies. Applying the four ICs to the remaining \textbf{\pubstepfive} studies  results in \textbf{\pubstepsix} included studies and \textbf{\pubstepsixout} excluded studies (\step{6}).}

\subsubsection{Selected Primary Studies}
\label{sec:selection-result}

\noindent
\change{The \textbf{\pubstepthreenew} primary studies obtained by the keyword-based query and the \textbf{\pubstepsix} primary studies obtained by snowballing result in a total of \textbf{\pubstepsixremainingnew} primary studies.
Among these studies, there are approaches that focus on model abstraction and bounded model checking techniques (\cf~IC2). However, they are not primarily concerned with counterexample explanation and interpretation, and therefore excluded by adding a third exclusion criterion (\textbf{EC3}): \textit{The primary study focuses mainly on model abstraction and bounded model checking techniques}.
Applying \textbf{EC3} to the \textbf{\pubstepsixremaining} studies excludes \textbf{\pubstepsevenout} primary studies and includes \textbf{\pubfinal} primary studies (\step{7}).
Overall, we select a corpus of \textbf{\pubfinal{}} primary studies for our survey.}

\subsubsection{Data Extraction}
\label{sec:extraction}

\noindent
The primary studies collected during the two phases are maintained in a shared spreadsheet to provide accessibility to all authors. 
To answer the research questions, we extract 20 data items from the primary studies.\footnote{The data items and their descriptions are listed in the appendix of this paper~\tabdataitems.} The extracted data is also maintained in the spreadsheet.

When answering the research questions, we present quantitative analysis results wherever reasonable and provide additional qualitative results and highlight exemplary publications.

%% file: tex/Analysis.tex
\section{Results}
\label{sec:analysis}

\noindent
This section provides the results of our study and answers the research questions introduced in \sect{RQ}. For quantitative RQs, we collect the answers and provide statistical data. For qualitative RQs, we extract relevant, interesting, or representative statements from the primary studies. The primary studies used for each research question are tabulated in \cite{appendix}.

\input{tex/rsq_representation.tex}
\input{tex/rsq_processed.tex}
\input{tex/rsq_enrich.tex}
\input{tex/rsq_input_domain.tex}
\input{tex/rsq_system_spec.tex}
\input{tex/rsq_tools.tex}
\input{tex/rsq_evaluation.tex}

%% file: tex/rsq_representation.tex
\subsection{\rsq{1} How are counterexamples explained and what are the effects of this explanation on counterexample interpretation?}
\label{sec:analysis:representation}

\begin{figure}
\begin{subfigure}{\linewidth}
	\includegraphics[width=\textwidth]{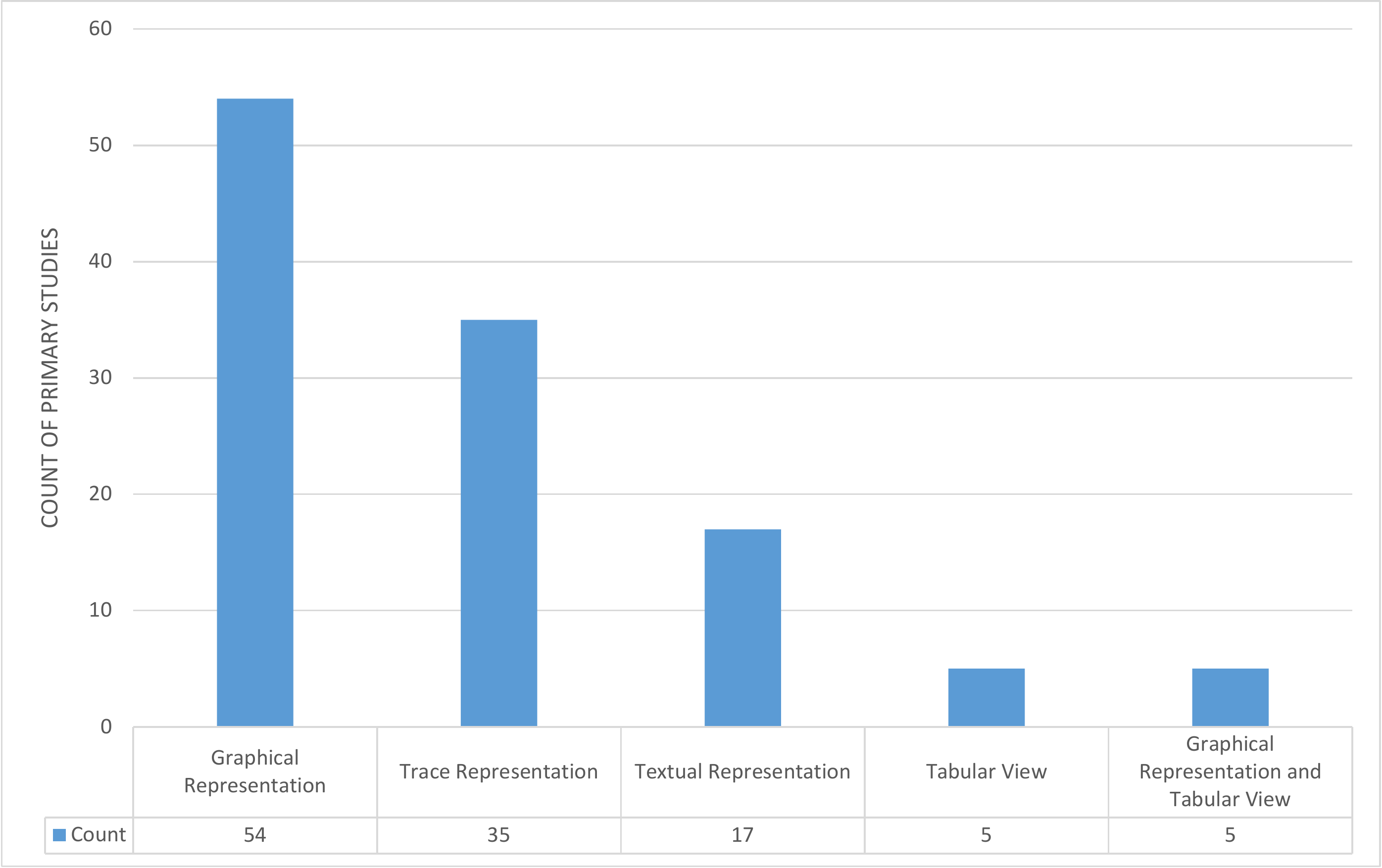}
	\caption{\rsq{1} Types of counterexample representation.}
	\label{fig:representation}
\end{subfigure}

\begin{subfigure}\linewidth
	\includegraphics[width=\textwidth]{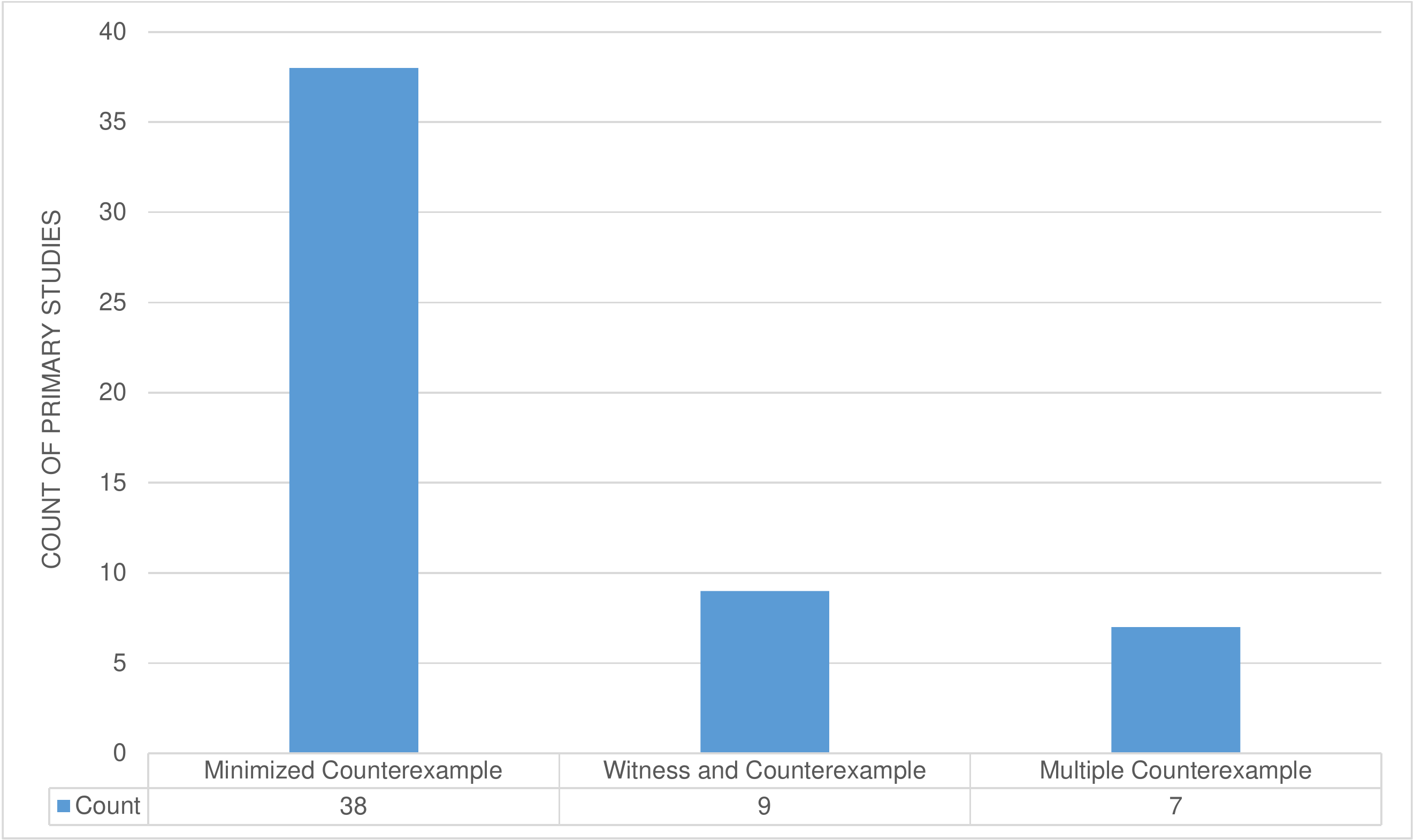}
	\caption{\rsq{2} Types of processed counterexamples.}
	\label{fig:processed}
\end{subfigure}
\caption{Types of counterexample representation and processed counterexamples.}
\end{figure}

\noindent
With this research question we investigate the counterexample explanations presented to engineers, particularly the style of representing a counterexample and what effects these have on counterexample interpretation.
The surveyed primary studies use five different styles of representation to explain counterexamples (\fig{representation}):
54 primary studies (47\% of all primary studies) use a graphical,
35 (30\%) a trace,
17 (15\%) a textual,
5 (4\%) a tabular representation, while the remaining
5 (4\%) use a combination of graphical and tabular representations.\footnote{The primary studies for each style of representation are listed in \tabrqone.}
In the following, we discuss each category with examples from the primary studies. We further provide qualitative statements from these studies to show the effects of counterexample explanation on the interpretation.

\subsubsection{Graphical representation}
\label{sec:analysis:representation:graphical}

\noindent
\emph{Graphical representation}s visualize a counterexample in a graphical notation, \eg as a state machine or block diagram.
Nguyen and Ogata~\cite{NguyenO17} state that \blockquote{a graphical representation of a counterexample would help human users comprehend it more intuitively}. Accordingly, we found \represgraphical primary studies (47\% of all primary studies) that use graphical explanations.
In addition, graphical \emph{animations} promise to further ease debugging of a counterexample~\cite{NguyenO17,Liu16a,LiL16}. 

\gls{UML} or \gls{SysML} graphical diagrams such as component, state-machine and sequence diagrams are used to explain counterexamples in \represfuml primary studies (50\% of the primary studies that use a graphical representation). Elamkulam \etal~\cite{ElamkulamGRKGKDM06} map counterexamples to states and events in the design model. These are presented as a \gls{UML} \emph{sequence diagram}, making it easy for the user to understand the design flaw. In addition, animation of the counterexample in graphical diagrams like \emph{function, block}~\cite{PakonenMLK13,PakonenTHP17,PakonenB17,PakonenBV18}, \emph{component}~\cite{RatiuGS19}, or \emph{state-machine diagram}~\cite{CampetelliJBDKZ15,Nguyen017-1} enables the user to simulate the counterexample step-by-step in the user-given input models to identify the error behavior.

\emph{Fault trees} are significantly smaller and easier to understand than corresponding stochastic counterexamples. A fault tree still contains all the information required to recognize the cause for the occurrence of a hazard~\cite{KuntzLL11}. Fault trees are used by \represfta primary studies to explain counterexamples (13\%).
For instance, the fault tree generation approaches by Leitner-Fischer and Leue~\cite{Leitner-FischerL13-1,Leitner-FischerL13-3,Leitner-FischerL14} compute causal events by performing causality checks for probabilistic models.
Notably, Leitner-Fischer and Leue~\cite{Leitner-FischerL13-1} claim from a case study that a \blockquote{fault tree is a compact and concise visualization of the counterexample, which allows for easy identification of basic events that cause the hazard}.
Similarly, Nguyen and Ogata~\cite{NguyenO17} argue that \enquote{\textit{it would be easier to comprehend a shorter counterexample, but the text representation of a counterexample need not be necessarily understandable. A graphical representation of a counterexample would help human users to comprehend it more intuitively}}.

\subsubsection{Trace representation}
\label{sec:analysis:representation:trace}

\noindent
A \emph{trace representation} is a modified form of a counterexample with addition or removal of (sub)-traces. Traces are the second most used type of representation in the primary studies (30\% of all primary studies). Traces can be used to represent a witness~(correct trace)~\cite{PeledPZ01}, a minimized counterexample~\cite{GastinMZ04}, or multiple counterexamples~\cite{ChechikG07}. A trace representation aims to lead to easier debugging and faster error comprehension. It is true for a minimized counterexample, the shorter the counterexample the simpler the debugging process~\cite{ChangBM07}. Further details of trace representations are discussed in \sect{analysis:processed}.

\subsubsection{Textual representation}
\label{sec:analysis:representation:textual}

\noindent
\emph{Textual representations} can be understood easily even by non-experts~\cite{BergSJ07} in formal methods. However, only \represtectual of the primary studies (15\% of all primary studies) use such a representation to provide statements for error notification or highlighting the error in a provided textual input system such as \gls{CNL}~\cite{LutebergetCJS17,AngelovCS13}, 
Structured English~\cite{FengGCT18,LutebergetJ18,CrapoMMR17,MoitraSCCDLYMM18,MoitraSCDLMMM19}, 
or error localization as known from programming languages~\cite{GroceKL04,BallNR03,ClarkeKL04,GroceCKS06,PuZ08}.

Feng \etal~\cite{FengGCT18} and Luteberget and Johansen~\cite{LutebergetJ18} generate a minimal set of structured language sentences. These \emph{natural language like} sentences have the potential advantage of providing diagnostic feedback to humans.
Berg \etal~\cite{BergSJ07} investigate two approaches for interpreting counterexamples: animation and natural language. Animation is a highly effective technique. However, animating and interpreting the railway interlocking and train behavior from the counterexamples is not always intuitive. For instance, \blockquote{the animation would show two trains using the same route and occupying the same segment at the same time. The user may incorrectly think there is an error as soon as this happens, however this is not the case.} For these reasons, authors deemed animation unsuitable for this use case and chose natural language for presentation instead.

\subsubsection{Tabular representation in combination with graphical representations}
\label{sec:analysis:representation:tabular}

\noindent
A \emph{tabular representation} is the least used type of counterexample representation by the primary studies (4\% of all primary studies). The \emph{combination of tabular and graphical representations} is used by additional five studies (4\% of all primary studies) to represent counterexamples. Tabular representations convert the variables and values of a counterexample into rows and columns without altering its original information. 

Frameworks like \gls{FASTEN}~\cite{RatiuGS19} and AutoFocus~3~\cite{HolzlF07} use tabular views to enhance readability. These frameworks transform a counterexample generated by \gls{NuSMV} into rows and columns, where the columns are the variables and rows represent the states. Arcaini \etal~\cite{ArcainiGR17a} as well as Bolton and Bass~\cite{BoltonB10} highlight the variable and its values in the table to show changes from the respective previous state.
Arcaini \etal~\cite{ArcainiGR17a} state that \blockquote{such representation is much more readable than the standard counterexample representation of \gls{NuSMV} in which only the variables that changed their value are shown}.
However, according to Loer \etal~\cite{LoerH06}, tables are primarily suitable for mechanical analysis, but hard to read by human analysts.

\subsubsection{Answer to RQ1}
\label{sec:analysis:representation:conclusion}

\noindent
The results of our study clearly show that graphical representations are used predominantly for the explanation of counterexamples. A format close to graphical representations, traces, is the second most used representation type, together representing 77\% of the surveyed primary studies.
Graphical representations are often based on component, state-machine and sequence diagrams, \eg from \gls{UML} or \gls{SysML}, as we observed for 50\% of the primary studies that use graphical representations. 
Concerning the effects of the explanation on the interpretation, primary studies prefer graphical or textual (natural language-like) explanations since they are easier to interpret than the counterexample or respectively an animated counterexample~\cite{NguyenO17,BergSJ07}.

%% file: tex/rsq_processed.tex
\subsection{\rsq{2} How are counterexamples processed and what effect does it have on interpreting the counterexample?}
\label{sec:analysis:processed}

\noindent
We investigate how counterexamples returned by a model checker are further processed to make them more valuable to the user. Out of the 116 primary studies, 54 (47\%) further process the counterexample while the remaining 62 studies (53\%) skip the processing and focus directly on representing the counterexample (\cf~\fig{overview}).

In our survey we found three main categories of \concept{processing a counterexample}, see \fig{processed}:
\processminim~primary studies (70\% of all primary studies that process the counterexample) process a counterexample to a minimized counterexample, \processwitness (17\%) to a witnesses, and \processmultip (13\%) to multiple counterexamples.\footnote{The primary studies for each of these categories are listed in \tabrqtwo.}
In the following, we discuss each category with examples from the primary studies. We further provide qualitative statements from these studies to show the effects of counterexample processing on the interpretation.
The interpretation is also influenced by the representation of the processed counterexample. Out of the 54 primary studies that process the counterexample, 35 (65\%) represent the processed counterexample as a trace, and the remaining 19 primary studies (35\%) explain the processed counterexample in a graphical or textual format.

\subsubsection{Minimized counterexample}
\label{sec:analysis:processed:minimized}

\noindent
\change{In this survey, we use the term \emph{minimized counterexample} for what is called \emph{shortest counterexample, abstract counterexample, reduced counterexample} in the surveyed primary studies, or is sometimes termed categorizing the error trace, transition, state, or variable in the counterexample. Schuppan and Biere~\cite{SchuppanB05} state that \blockquote{most counterexamples still need to be interpreted by humans, and shorter counterexamples will, in general, be easier to understand}.
This is often done by highlighting the erroneous transitions to distinguish them from correct transitions.} \processminim~primary studies (70\% of all primary studies that process a counterexample) follow such an approach, and can be grouped into three categories based on the kind of specification and verification model: qualitative, real-time, and probabilistic. 
Notably, we found only one primary study~\cite{GerkingSDH15} that uses a real-time specification and minimizes the counterexample by removing variables that are irrelevant for error comprehension. The two other categories are discussed in the following.

\textbf{Counterexample minimization for qualitative specifications.}
27 of the \processminim primary studies (71\%) minimize a counterexample for qualitative systems.
From the surveyed primary studies, the most prominent model checkers used to perform counterexample minimization are \gls{NuSMV}~\cite{SchuppanB05,ShenQL05-2,WeitlN10,ShenQL05-3,HeljankoJKLL06}, \gls{SPIN}~\cite{GastinMZ04,GastinM07,EdelkampLL04,HansenG08,EdelkampLL01}, and Maude~\cite{NguyenO17}.
The first novel approach to minimize loop-like and path-like counterexamples for given ACTL properties is proposed by Shen \etal~\cite{ShenQL05-3}, while the existing approaches can only deal with path-like counterexamples of invariants of the form $AG f$.
One of the notable approaches we found in our survey is from Hansen and Geldenhuys~\cite{HansenG08}, where the user can specify the size of the shortest counterexample, which is shown to the user simultaneously during the verification run. This allows the user to interrupt the algorithm when the counterexample shrinks short enough. Moreover, the user can easily specify the limits on the number of states and transitions that the algorithm is allowed to explore or---more directly---the time and memory it is allowed to consume. 
\gls{BFL} is the most effective counterexample minimization algorithm~\cite{ShenQL05-1,ShenQL05-2}. It performs refutation analysis of counterexample variables, \ie to extract the set of variables that are irrelevant to the counterexample. Thus, multiple variables can be eliminated with only one call to the \gls{SAT} solver.

\textbf{Counterexample minimization for probabilistic specifications.}
Ten of the \processminim primary studies that minimize counterexamples (26\%) specifically address probabilistic systems. For probabilistic model checking, a set of counterexamples is typically required to provide an accumulated probability mass that violates the probability bound of the specified probabilistic property~\cite{Leitner-FischerL13-2}. Thus, to compute the probabilities for all traces, several traces need to be understood and analyzed. Therefore, Leitner-Fischer and Leue~\cite{Leitner-FischerL13-2} propose 
\emph{Causality Checking}~\cite{Leitner-FischerL13-1} that is integrated into the state-space exploration algorithm for qualitative model checking and that computes causality relationships on-the-fly. 
Consequently, the probability computation can be limited to the causal events. To apply the 
causality computation to a \gls{PRISM} model, the \gls{DiPro} is used to generate a counterexample.
Further, Debbi and Bourahla~\cite{DebbiB13-1} propose an algorithm for diagnosis generation along with a set of causes from the counterexample generated by \gls{DiPro}. It is the first approach that introduces the diagnosis for counterexamples in probabilistic model checking.

Abraham \etal~\cite{AbrahamJWKB10} generate an abstract counterexample by choosing one or more 
paths from a counterexample, which carry enough probability. The user can choose the level of abstraction of a counterexample and refine the counterexample step-by-step by choosing abstract nodes and replacing them by the concrete input nodes.
Similarly, the \gls{GUI} presented by Jansen \etal~\cite{JansenAVWKB12} generate concrete and abstract graphs for \gls{DTMC} that can be stored, loaded, abstracted, and concretized by the user, thus controlling the hierarchy of a counterexample.
The interactive visual functions used by  Aljazzar and Leue~\cite{AljazzarL08} allow users to selectively filter the displayed information, which helps to focus during visual analysis.

\paragraph{Methods for counterexample minimization}

Out of 38\,primary studies that minimize the counterexample, 15 (39\%) follow a specific or adapt an existing algorithm such as \gls{BFL}~\cite{RaviS04,ShenQL05-1,ShenQL05-2} to perform the minimization. 
In the following, we discuss other commonly found methods to minimize counterexamples that are either based on search, translation and abstraction, or comparison of the counterexample with the correct system behavior.

\textbf{Search.}
\label{sec:analysis:method:minimization:search}
Among the \processminim\,primary studies that minimize counterexamples, \represfsearch (37\%) use search methods for this purpose. In general, a search method explores a given system state space in a directed way by representing the system behavior in a graph structure, in which states are nodes and transition are edges. For example, Tan \etal~\cite{TanACZL04} reduce the size of the counterexample using a heuristic-guided search strategy. Every node is associated with a value denoting the distance to the non-accepting state that does not have any outgoing edge except of self-loops. During the search, the shortest path is computed based on this distance and returned as a minimized counterexample.

Edelkamp \etal~\cite{EdelkampLL01,EdelkampLL04} as well as Aljazzar and Leue~\cite{AljazzarL08} use the directed search algorithm \emph{A$^\ast$} that follows a \gls{DFS} method to find the shortest counterexample~\cite{EdelkampR98}.
Edelkamp \etal~\cite{EdelkampLL01,EdelkampLL04} further propose a heuristic that accelerates the search in the direction of a specified failure situation, with an improved nested \gls{DFS} to identify a safety property violation.  
Aljazzar and Leue~\cite{AljazzarL08} use A$^\ast$ to find the shortest path in a directed graph. First, the state space of a Markov chain is represented as a directed graph, in which nodes represent states and edges represent transitions. Then, A$^\ast$ explores all edges of the graph and inserts them into a 
so-called \emph{path graph}. With the generated path graph, an interactive visualization allows filtering out parts of the state space.

Gastin and Moro~\cite{GastinM07} state that nested \gls{DFS} approaches strongly rely on the order of the transitions.
Therefore, they propose a \emph{polynomial-time algorithm} based on \gls{BFS} where the ordering of the transitions has no impact. The proposed algorithm computes a counterexample of minimal size for \gls{SPIN} that uses forward edges and consumes less memory than SPIN while trying to reduce the size of counterexamples.
Similarly, Hansen and Kervinen~\cite{HansenK06} present an algorithm based on \gls{BFS} that finds a minimal counterexample.
Furthermore, the model checker Maude~\cite{NguyenO17} is equipped with a search command that exhaustively traverses the reachable states in \gls{BFS} manner to generate the shortest counterexample.
Aljazzar and Leue~\cite{AljazzarL10} use an \gls{XBF} algorithms that is guided by heuristics to amplify the power of search in terms of computational efforts and counterexample quality. The algorithm explores the state transition graph of \glspl{CTMC} and \glspl{DTMC}, and  incrementally determines a sub-graph of the state transition graph, which covers the most probable diagnostic paths. 

Apart from using either a \gls{BFS} or \gls{DFS} search method, Leitner-Fischer and Leue~\cite{Leitner-FischerL14} integrate \gls{DFS} and \gls{BFS} for the causality check.
The resulting search algorithm distinguishes between bad and good executions 
based on which the causality relationships are computed.
Similarly, the algorithm by Groce and Visser~\cite{GroceV03} generates a subset of \emph{negatives}---executions that reach a particular error state---and a set of potential \emph{positives}---any execution that does not end up in the error state. Then, it uses a model checker to explore backward from the original counterexample. The algorithm is depth-first, but it can be integrated into both breadth-first and heuristic-based model checking algorithms.

\textbf{Translation and abstraction.}
\label{sec:analysis:method:minimization:abstraction}
From the \processminim\,primary studies that minimize counterexamples, 5\,studies (13\%) use a method based on translation and abstraction of the modeled system or subsystem for this purpose.
Frameworks as proposed by Gerking \etal~\cite{GerkingSDH15}, as well as \gls{FASTEN}~\cite{RatiuGS19} and AutoFOCUS~3~\cite{KanavA17} allow engineers to model a system with \glspl{DSL}. 
In these frameworks, a \gls{DSL} supports to minimize the counterexample by performing two types of translation: forward and backward translation.
The forward translation adds new states and transitions to the verification model to enable model checking when translating the design model expressed in a \gls{DSL} to the verification model. The backward translation removes the states and transitions being added by the forward translation, thus reducing the size of the counterexample. Consequently, the complexity of counterexample interpretation is reduced when lifting the counterexample to the \gls{DSL}-based design models.

Jansen \etal~\cite{JansenAKWKB11} compute critical subsystems hierarchically by concretizing abstract states and reducing the concretized parts. Such subsystems reduce the number of involved states and transitions when  generating a counterexample. The computation of critical subsystems is based on finding most probable paths or path fragments to be contained in the critical subsystems. Concretization of only the user-relevant parts of the abstract critical subsystems allows for an intuitive approach to error correction. 
To compute state-minimal critical subsystems for \glspl{DTMC} and \glspl{MDP}, Wimmer \etal~\cite{WimmerJABK12} propose \gls{SMT} and \gls{MILP}-based formulations. Wimmer \etal~\cite{WimmerJAKB14} further propose a tool that yields 
smaller subsystems than the available heuristic-based tools, with a lower bound for the model checker to provide a minimal counterexample.

\textbf{Comparison with correct system behavior.}
\label{sec:analysis:method:minimization:comparision}
Out of the \processminim\,primary studies that minimize counterexamples, 4 studies (11\%) identify erroneous and correct states in the counterexample by comparing the counterexample with the correct system behavior.
Barbon \etal~\cite{Barbon0SY19,BarbonLS19,BarbonLS18,Barbon0S18} propose an approach to simplify the understanding of counterexamples for a \gls{LTS}. The approach extracts actions from the counterexample that are relevant for the violation to reduce the amount of debugging information. To do this, it compares the generated counterexamples with the correct behaviors of the verification model, and differentiates the error transitions and correct transitions in the counterexample. Similarly, the approach by Kaleeswaran \etal~\cite{KaleeswaranNVG20} identifies erroneous and correct states in the counterexample. It particularly supports engineers in locating faults in a Contract-Based Design (CBD), a user-provided design model, that does not pass the refinement check.

\subsubsection{Witness trace}
\label{sec:analysis:processed:winess}

\noindent
\processwitness primary studies (17\% of all 54 studies that process a counterexample) generate \emph{witness traces} from counterexamples. In addition to counterexamples, which show traces that fail to satisfy specifications, witnesses show traces that can satisfy the failed specification. The user can compare witness traces with the counterexample to understand the error behavior and propagation. As two prominent approaches, Peled \etal~\cite{PeledPZ01} generate a deductive proof so that the system meets its \gls{LTL} specification and Groce \etal~\cite{GroceCKS06} produce the successful execution that is most similar to the counterexample.

Satisfying a specification, a \emph{witness trace} is the opposite view of the counterexample. 
Peled \etal~\cite{PeledPZ01} state that \blockquote{\emph{proof by lack of counterexample} is the main drawback of the model checking approach; some would even say that model checking is a tool for falsification rather than a tool for verification}. Thus, their work generates a deductive proof by using the \gls{LTL} model checking process and presents it as a graph when the system meets its \gls{LTL} specification. Therefore, it is possible to justify why the system actually works. Similarly shown by Beyer \etal~\cite{BeyerDDHS15}, the two-step approach provides evidence to support this claim, \ie whether it produces a witness for correctness or violation of the specification.

Jin \etal~\cite{JinRS04} present an enhanced error trace as an alternative to fated (forced) and free segments. The fated segments show unavoidable progress towards the error while the free segments show the choices that, when avoided, might have prevented the error. Hence, the demarcation into segments tends to highlight critical events. Similarly, Groce and Visser~\cite{GroceV03}  generates \emph{negatives}---executions that reach a particular error state---and \emph{positives}---any executions not ending in the error state---to find a counterexample by searching close to a path the user suspects could lead to an error.

Leue and Befrouei~\cite{LeueB12} use \gls{SPIN} to detect deadlocks in concurrent systems, where a set of counterexamples is generated as a ``bad'' dataset and an additional 
``good'' dataset is generated that does not violate the specification. By comparing the bad with the good dataset, sequences of actions are extracted to locate the cause of the occurrence of a deadlock. Two different approaches are used by Kumar \etal~\cite{KumarKV05} for error explanation. The first approach is to compute the minimal change in a system that eliminates the error causing the counterexample. The second approach computes a correct trace of the system that is close to the counterexample.

\subsubsection{Multiple counterexamples}
\label{sec:analysis:processed:multiple}

\noindent
Another technique we found in \processmultip primary studies (13\% of all primary studies that process a counterexample) provides \emph{multiple counterexamples}. This is usually the result of generating all possible counterexamples for every possible failure case, instead of providing just a single counterexample. Standard model checkers like \gls{NuSMV} do not produce multiple counterexample by default. Copty \etal~\cite{CoptyIWKK03} present a wizard that can retrieve multiple counterexamples in a single run and can be used to analyze all possible error conditions for the provided specification. PyNuSMV generates multiple or complete counterexamples for branching logics like \gls{CTL} by translating specifications into $\mu$-calculus~\cite{BusardP18}.

The method of multiple counterexamples enables finding all possible failures in a single run of verification. It also has the ability to identify more than one root cause. This can reduce the number of verification runs greatly~\cite{CoptyIWKK03}.
Dominguez and Day~\cite{DominguezD13} provide an approach that produces all counterexamples automatically by modifying the specification and grouping counterexamples together. Not modifying the model checker engine or the verification model, this approach works with any \gls{LTL} model checker, and produces the complete set of counterexamples. Ball \etal~\cite{BallNR03} propose a technique to localizes the error cause in the error trace and generates multiple error traces.
This technique generates multiple error traces by making use of the existing correct traces to localize the error cause. This is similar to SLAM~\cite{BallR02} that verifies C programs and localizes the errors in device drivers. The algorithm performs the following functions in the program: highlighting each error location, introducing halt statements, and re-running the model checker to produce additional error traces.
Finally, the framework by Chechik and Gurfinkel~\cite{ChechikG07} provides a simple and unified way to interact with the counterexample generator, which combines property-based and model-based choices.

\subsubsection{Answer to RQ2}
\label{sec:analysis:processed:conclusion}

\noindent
Counterexample minimization is by far the dominant method for processing counterexamples in the context of counterexample explanation. \processminim primary studies use this method (33\% of all primary studies and 70\% of all studies that process a counterexample). \change{This strengthens our hypothesis that highlighting all the essential information like erroneous states and variables in the concrete counterexample is helpful for non-experts for error comprehension.}

The model checkers \gls{SPIN} and Maude are most often used for applying search methods such as \gls{DFS} and \gls{BFS} to minimize a counterexample.
Model transformation is another method used for minimizing counterexamples. It relies on domain-specific design models from the user and transform them to formal verification models. Further, the counterexample generated during verification is translated back to the design models provided by the user.
This hides the complexity of using a formal notation and logic and thus, the approach performs both the model abstraction and the counterexample representation.

One of the biggest advantages of model checking is the generation of a counterexample for error identification. Contrary to counterexamples, witnesses provide traces that satisfy the specification. This enables  users to identify quickly the offending behavior and supports debugging by comparing the correct with the offending behavior.
\change{As per Copty \etal~\cite{CoptyIWKK03}, one of the core advantages of multiple counterexamples is that all the possible counterexamples are identified in a single run which, enables a user to find all possible error root causes.} 

%% file: tex/rsq_enrich.tex
\subsection{\rsq{3} What kind of additional information is used to enrich the counterexample explanation?}
\label{sec:analysis:enrich}

\noindent
According to Pakonen \etal~\cite{PakonenBV18}, \blockquote{visualization of each step of the counterexample trace in the model may not be sufficient for fast understanding of the essence of the counterexample}. This statement motivates the use of additional information to enrich counterexample explanations to ease error comprehension.
However, in our survey we only found 8 primary studies (7\% of all primary studies) that provide additional information along with the counterexample explanation.\footnote{These primary studies are listed in \tabrqthree.} 
Consequently, we collected qualitative statements from the primary studies that illustrate such additional information, grouped by the type of counterexample representation.

\subsubsection{Information enriching graphical representations of counterexamples}
\label{sec:analysis:enrich:graphical}

\noindent
We only found 3 primary studies (6\% of all 54 primary studies that use graphical representations of counterexamples) that enrich graphical representations of counterexamples with additional information.
A cross-platform tool by Pakonen \etal~\cite{PakonenBV18} visualizes the counterexample generated for \gls{LTL} specifications by \gls{NuSMV} by highlighting the values of atomic propositions, which are important for understanding the counterexample. 
Similarly, Beer \etal~\cite{BeerBCOT12} find a set of causes for a specification failure on the given counterexample trace, using the notion of causality introduced by Halpern and Pearl~\cite{HalpernP2000}. These causes are presented to the user as a visual notification.

Error localization is not only performed for the counterexample or specification, but also for the design or verification model. \gls{STANCE} builds the complete counterexample using the Simulink simulator and analyzes this counterexample to compute the causes~\cite{BochotVWW10}. The resulting colored model allows the user to focus on the structural part of the model that is responsible for violating the property.

\subsubsection{Information enriching textual representations of counterexamples}
\label{sec:analysis:enrich:textual}

\noindent
5 primary studies (29\% of all primary studies that use textual representations of counterexamples) provide additional information that enrich textual representations of counterexamples. Particularly, domain ontologies and vocabularies are used to enhance the counterexample with additional information to ease error comprehension. Crapo \etal~\cite{CrapoM19,CrapoMMR17} and Moitra \etal~\cite{MoitraSCCDLYMM18,MoitraSCDLMMM19} use \gls{RAE} and \gls{ASSERT} that accept a formal requirement in an easily understandable syntax by making use of a domain ontology. Further, they analyze an incomplete set of requirements and localizes the error by identifying the responsible requirements with an error marker. Errors, warnings, and informational markers are used to make the user aware of the errors and ambiguities. Likewise, Berg \etal~\cite{BergSJ07} use a domain-specific vocabulary to present errors in railway interlocking systems in a language similar to natural language. They further present a table explaining an action of the railway interlocking systems relevant to the error, and the corresponding action that permits the error to occur. 

\subsubsection{Answer to RQ3}
\label{sec:analysis:enrich:conclusion}

\noindent
Error localization is used to highlight or to identify the cause of an error in a given design or verification model. For example, Pakonen \etal~\cite{PakonenBV18} highlight atomic proposition values, Beer \etal~\cite{BeerBCOT12} mark errors in a given specification, and \gls{ASSERT}~\cite{MoitraSCCDLYMM18,MoitraSCDLMMM19} highlights the responsible requirements with error markers. These approaches highlight errors in the given input design model, while the approach by Berg \etal~\cite{BergSJ07} renders an explicit explanation, further aiming to ease error comprehension.

%% file: tex/rsq_input_domain.tex
\subsection{\rsq{4} For which input domains are the counterexample explanation approaches available? What is the influence of the input domain on the explanation?}
\label{sec:analysis:domain}

\noindent
To answer this research question, we collected the input and output domains used by the primary studies (\cf~\fig{overview}). 
We investigate whether counterexamples are represented and explained within the input domain or whether an output domain different from the input domain is used. Moreover, we collected qualitative data that demonstrate the impact of the used domains on the explanations of counterexamples.
For this research question, we excluded 35 primary studies that use trace representation, since these keep the original format of the counterexamples. Thus, we investigated the remaining 81 primary studies that transform counterexamples to graphical, textual, or tabular representations.

\subsubsection{Input and output domains in counterexample explanation}
\label{sec:analysis:domain:ip_op}

\begin{figure}
\begin{subfigure}{\linewidth}
	\includegraphics[width=\textwidth]{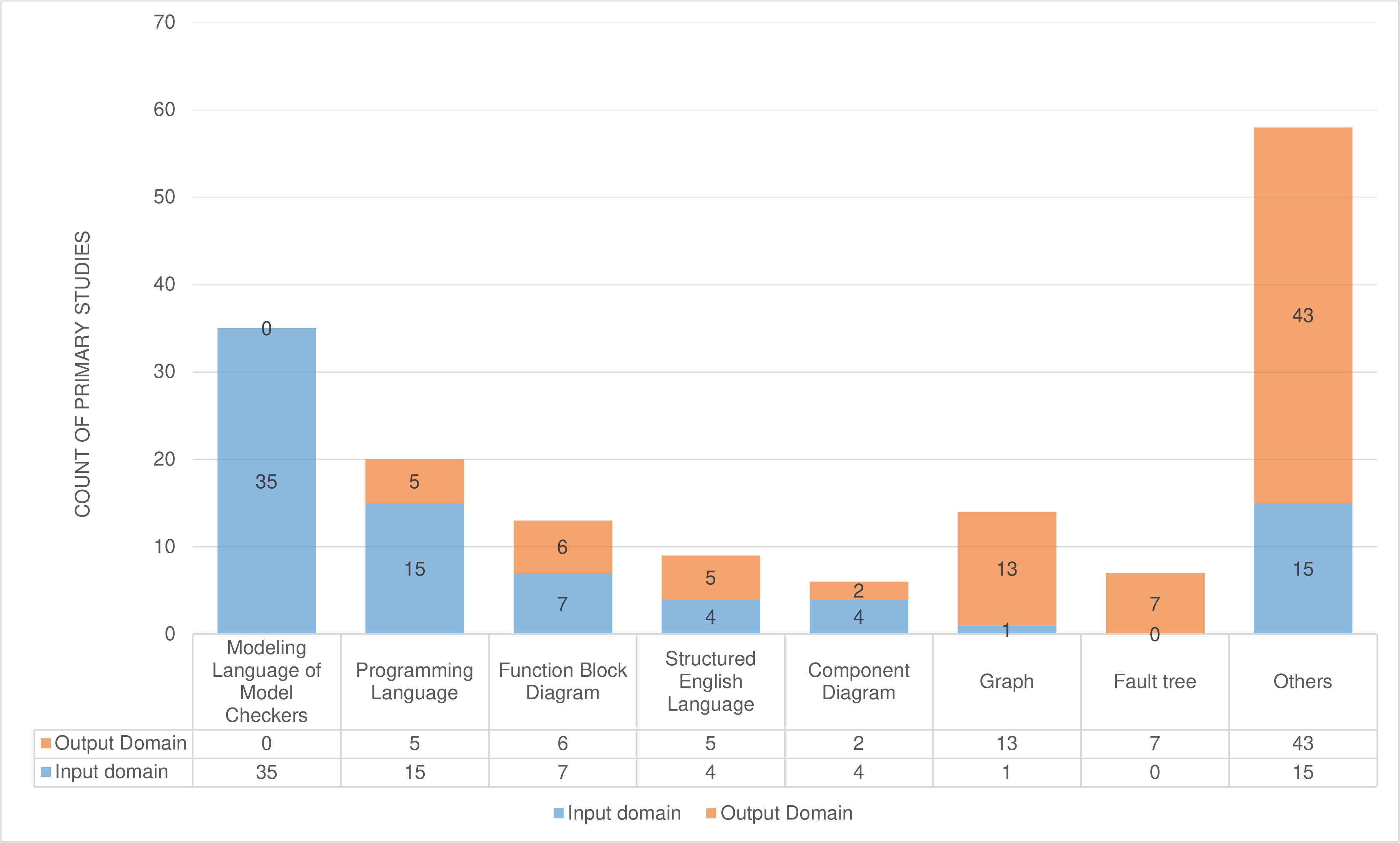}
	\subcaption{Input (blue) and output (orange) domains used by counterexample explanation approaches. Domains that are used as input or output in less than six primary studies are counted as \enquote{Others}. These other domains are CNL, SCADE/Simulink model, Control signal table, Requirement, UAV mission planning, GPS Image, BPMN Model, Graph, LSC, RTL Model, UML are input domains. Property, MSC, UAV mission planning, GPS Image, Sequence Diagram, MSC, BPMN Model, LSC and STD, Sequence Diagram, Flow Chart, State machine, and Tabular View are output domains.}
	\label{fig:input_output-domains}
\end{subfigure}

\begin{subfigure}{\linewidth}
	\includegraphics[width=\textwidth]{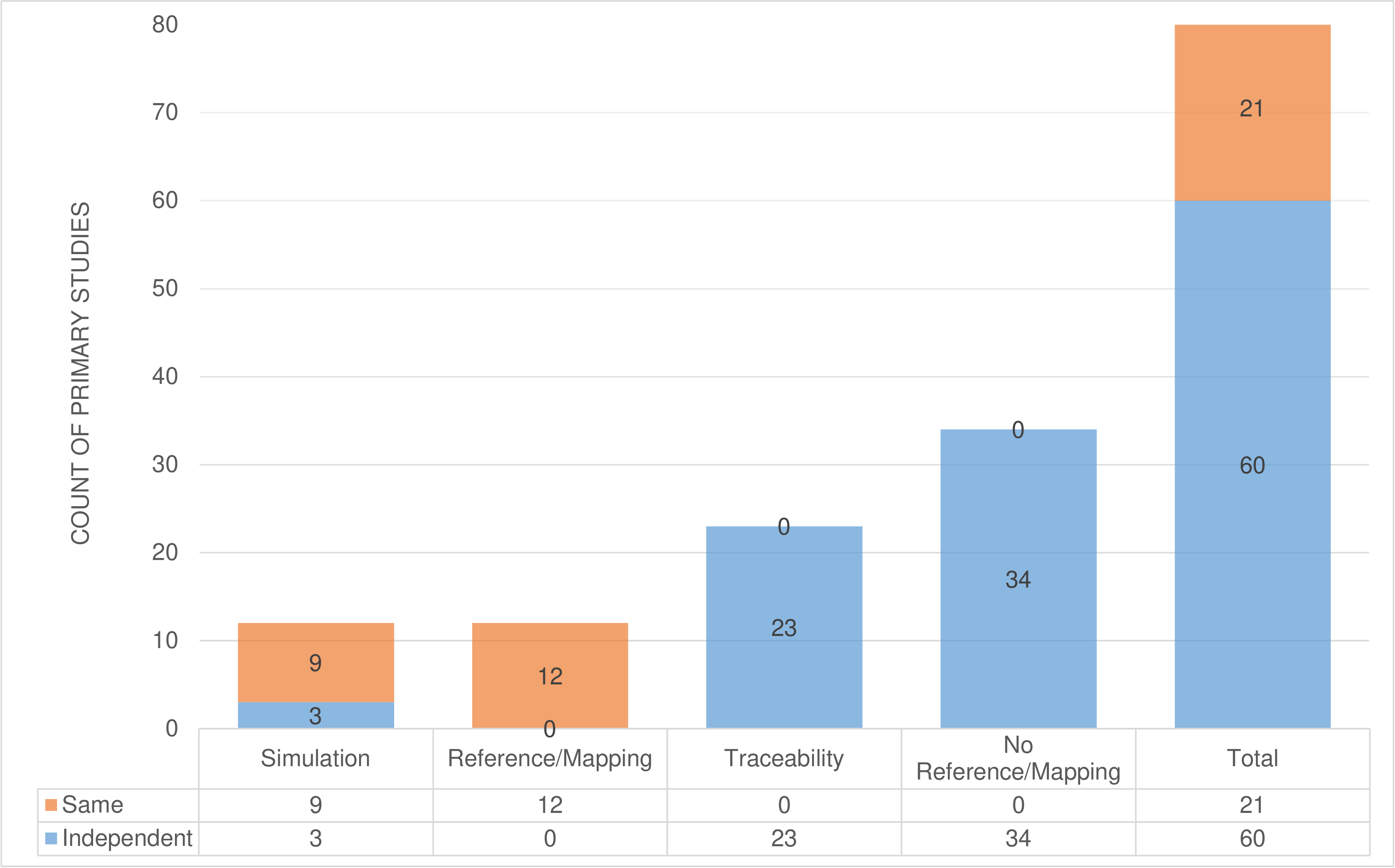}
	\subcaption{Influences of the input domain on the counterexample explanation.}
	\label{fig:relinout}
\end{subfigure}
\caption{\rsq{4} Input and output domains, and influences of the input domain on the counterexample explanation.}
\end{figure}

\noindent
\fig{input_output-domains} shows the count of primary studies that use certain input (blue bars) and output domains (orange bars).\footnote{The primary studies for each input and output domain are listed in \tabrqfourab.}
In the following, we discuss each domain with examples from the primary studies. We further provide qualitative statements from these studies to show the effects of the domain on counterexample explanation and interpretation.
	
The most used input domain is the \emph{modeling language of the used model checker} that we found in \inmlmc primary studies (30\% of all primary studies).
The second-most used input domain are \emph{programming languages} such as C, ANSI-C, or \gls{PLC} (\inprogram primary studies, 13\% of all primary studies). Among these \inprogram studies, \outprogram studies
explain the counterexample in the given input programming language. For example, the approach by Clarke \etal~\cite{ClarkeKL04} verifies ANSI-C programs, and the error is localized right in the given input program. The graphical interface presents the counterexample traces and allows stepping through the trace in the same way as a debugger allows stepping through a program. 

Pakonen and Bj{\"o}rkman~\cite{PakonenB17} state in the context of counterexample explanation that \blockquote{one of the strengths of function block diagrams as a programming language is that it is relatively easy to understand the flow of processing from the inputs to the outputs}.
Accordingly, we found two component-based input and output domains in our survey: \emph{function block diagrams} and \emph{component diagrams}. Functional block diagrams are used by seven primary studies as an input domain and six primary studies as an output domain. Similarly, component diagrams are used by four primary studies as an input domain and two primary studies as an output domain. In total, 11 of all primary studies (9\%) use a component-based domain for the input, and 8 of all primary studies (7\%) use such a domain for the output.
Frameworks such as MODCHK, FASTEN, and AutoFOCUS~3 represent the input in a component-based domain.
MODCHK~\cite{PakonenBV18,PakonenTHP17,PakonenB17} further simulates the counterexample in the given input domain, a function block diagram. Similarly, the frameworks \gls{FASTEN} and AutoFOCUS~3 simulate counterexamples in the provided input component diagram. 

A natural language like input domain that we found in our survey is \emph{Structured English Language} (\insel primary studies, \ie 3\% of all primary studies), which overshadows the complexity of formal notations for engineers~\cite{CrapoMMR17,MoitraSCCDLYMM18}. The \gls{ASSERT} framework by Crapo \etal~\cite{CrapoMMR17} and Moitra \etal~\cite{MoitraSCCDLYMM18,MoitraSCDLMMM19}, as well as the RailComplete framework by Luteberget and Johansen~\cite{LutebergetJ18} accept Structured English Language as the input domain. 
To improve the interpretation of counterexamples, \outsel primary studies (4\% of all primary studies) use \emph{Structured English Language} as the output domain. One of the examples for generating Structured English Language referring to the counterexample is presented by Feng \etal~\cite{FengGCT18} who define a set of Structured English Language sentences to describe the requirement violation of robotic behavior. An example for such a structured sentence is the template \blockquote{The robot $<$action$>$ when $<$proposition$>$}~\cite{FengGCT18}. A violation of a requirement is explained using the template by referring to the counterexample and replacing $<$action$>$ with possible robotic actions, and the $<$proposition$>$ is replaced with (conjunctions of) violated atomic propositions that represent the robot's configuration.

\emph{Graphs} (\outgraph primary studies, \ie 11\% of all primary studies) and \emph{Fault trees} (\outft primary studies, \ie 6\% of all primary studies) are the most-used output domains for counterexample explanation in the primary studies, which does not contribute much for input domains. One of the frameworks that presents a counterexample as a graph is \gls{COMICS}~\cite{JansenAVWKB12}. It allows the user to visualize either the original graph or the abstract graph, translated from the counterexample. A fault tree provides a compact and concise representation of system failures using a graphical notation well known to safety engineers~\cite{KuntzLL11}. The work by Leitner-Fischer and Leue~\cite{Leitner-FischerL13-3,Leitner-FischerL14} is an example for using visual representations of fault trees as the output domain.
To achieve this, they perform a causality check that computes event combinations causing a property violation, together with the order in which the events have to occur to be causal.

\subsubsection{Relationship between input and output domains}
\label{sec:analysis:domain:inout}

\noindent
The input domain, particularly the way systems are modeled, influences the explanation of the counterexample using a certain output domain. We categorize relationship between the input and output domains mainly into two different groups: explanation of the counterexample is done in an output domain \textit{identical} to the input domain or \emph{different} from the input domain. An overview is given in \fig{relinout} that further details these two groups into four categories:
\begin{enumerate}\itemsep0em
	\item \textit{Reference/Mapping:} Variables from the counterexample and their static values are attached to the input model or referred to in the input model.
	\item \textit{Simulation:} In addition to reference/mapping, variables from the counterexample and their values are animated in the provided input model or in any of the graphical representation, \eg their dynamic values can be displayed for each step of the counterexample.
	\item \textit{Traceability:} The counterexample is explained independently of the input domain, but the input domain is referred to in the counterexample explanation. In this case, the input and output domains are different.
	\item \textit{No Reference/Mapping:} The input and output domains are completely independent of each other; the counterexample explanation does not have any reference to the input.
\end{enumerate}

\fig{relinout} shows that from 81 primary studies, 60 (74\%) use an output domain \emph{different} from the input domain. Only \inoutdep primary studies (26\%) provide explanations of counterexamples using an output domain \textit{identical} to the input domain.\footnote{The primary studies for each category are listed in \tabrqfourc.} 

\paragraph{Identical input and output domains}
\label{sec:analysis:domain:same}

\noindent
9 of the \inoutdep primary studies that use identical input and output domains (43\%) perform \textit{simulation}, and 12 (57\%) rely on \textit{reference/mapping}.
Simulation by step-wise animation in both forward and backward directions is one of the ways to display and explain a counterexample. \gls{FASTEN}~\cite{RatiuGS19} and AutoFOCUS\,3~\cite{KanavA17} use the component diagram and MODCHK~\cite{PakonenBV18,PakonenTHP17, PakonenB17,PakonenMLK13} uses a functional block diagram as the input and output domain. In these frameworks, a simulator animates the counterexample back and forth to display every state variable with its values. Similarly, Groce \etal~\cite{GroceKL04,GroceCKS06} present a graphical user interface designed for \gls{CBMC}~\cite{KroeningT14} that allows the users to interactively step through the counterexample traces, in the provided input programming language.

Another option to visualize counterexample traces in the input domain is by referencing/mapping it to the user-provided design model. \gls{STANCE}~\cite{BochotVWW10,CastillosWW15} has been developed in the Matlab/Simulink environment and the complete counterexample is visualized using a Simulink model where paths of the cause are colored. Muram \etal~\cite{MuramTZ15} present the visual support based on the information provided by the counterexample analyzer mapped to the input, a \gls{BPMN} model. 
Particularly, the model element that indicates the violation of an \gls{LTL} property, the containment violations, and the elements that satisfy the property are highlighted in different colors.

\paragraph{Different input and output domains}
\label{sec:analysis:domain:different}

\noindent
As shown in \fig{relinout}, 23 of the 60 primary studies with different input and output domains (38\%) use \emph{traceability}, 34 (57\%) use \emph{no reference/mapping}, and 3 (5\%) use \emph{simulation}. 
Considering this research question and especially the influence of the input domain on the counterexample explanation, we cannot determine any such influence for the category \emph{No reference/mapping} since the input and output domains do not have any relationship to each other.

Considering traceability, the approaches by Berg \etal~\cite{BergSJ07} and Feng \etal~\cite{FengGCT18} use domain terminology as a vocabulary to represent the counterexample in a natural language like format. Such approaches maintain traceability to access the vocabulary from the provided input in order to generate counterexample explanations.
Likewise, in the work by Angelov \etal~\cite{AngelovCS13}, the counterexample of \gls{CLAN} is explained in a \gls{CNL}. The work by Luteberget and Johansen~\cite{LutebergetJ18} represent the requirement violation as a textual message to the railway engineer, which contains a reference to the rule source.

Considering simulation, the counterexample representation based on state machines in AutoFOCUS~3~\cite{CampetelliHN11,CampetelliJBDKZ15} can simulate either a path from the initial state to an erroneous state or a loop. Each step of the counterexample sequence is described by the actual value of variables and the i/o ports of the user-provided design model. 

\subsubsection{Answer to RQ4}
\label{sec:analysis:domain:conclusion}

\noindent
Our results show that most counterexample explanation approaches use the language of the verification tool as the input domain, indicating that most approaches are targeted to tool experts, having a fair knowledge in formal methods.
Counterexamples represented as fault trees in programming languages and function block diagrams support the corresponding domain experts to understand the error without the need for deep understanding of the formalism used by the model checker. This is especially true if the counterexample is simulated or animated in the given input domain.
However, we only found a few primary studies for this technique~\cite{PakonenBV18,RatiuGS19,PakonenTHP17,PakonenMLK13,PakonenB17,KanavA17}.

In most of the primary studies, the output domain is different from the input domain. Approaches like simulation of counterexamples extract the required information from the input domain and represents the error information independently.
Graphical representations and textual languages like Structured English or \gls{CNL} usually refer to the input domain and use the input domain as the primary source for generating a counterexample explanation.

%% file: tex/rsq_system_spec.tex
\subsection{\rsq{5} What are the different temporal logics used to express system specifications and what type of properties are covered in counterexample explanation approaches?}
\label{sec:analysis:spec}

\noindent
To answer this research question, we provide quantitative results of the temporal logics and types of properties that are used by the primary studies. 

\subsubsection{Specification logic}
\label{sec:analysis:spec:formalism}

\begin{figure}
	\begin{subfigure}[t]{0.6\linewidth} 
	\includegraphics[width=\textwidth]{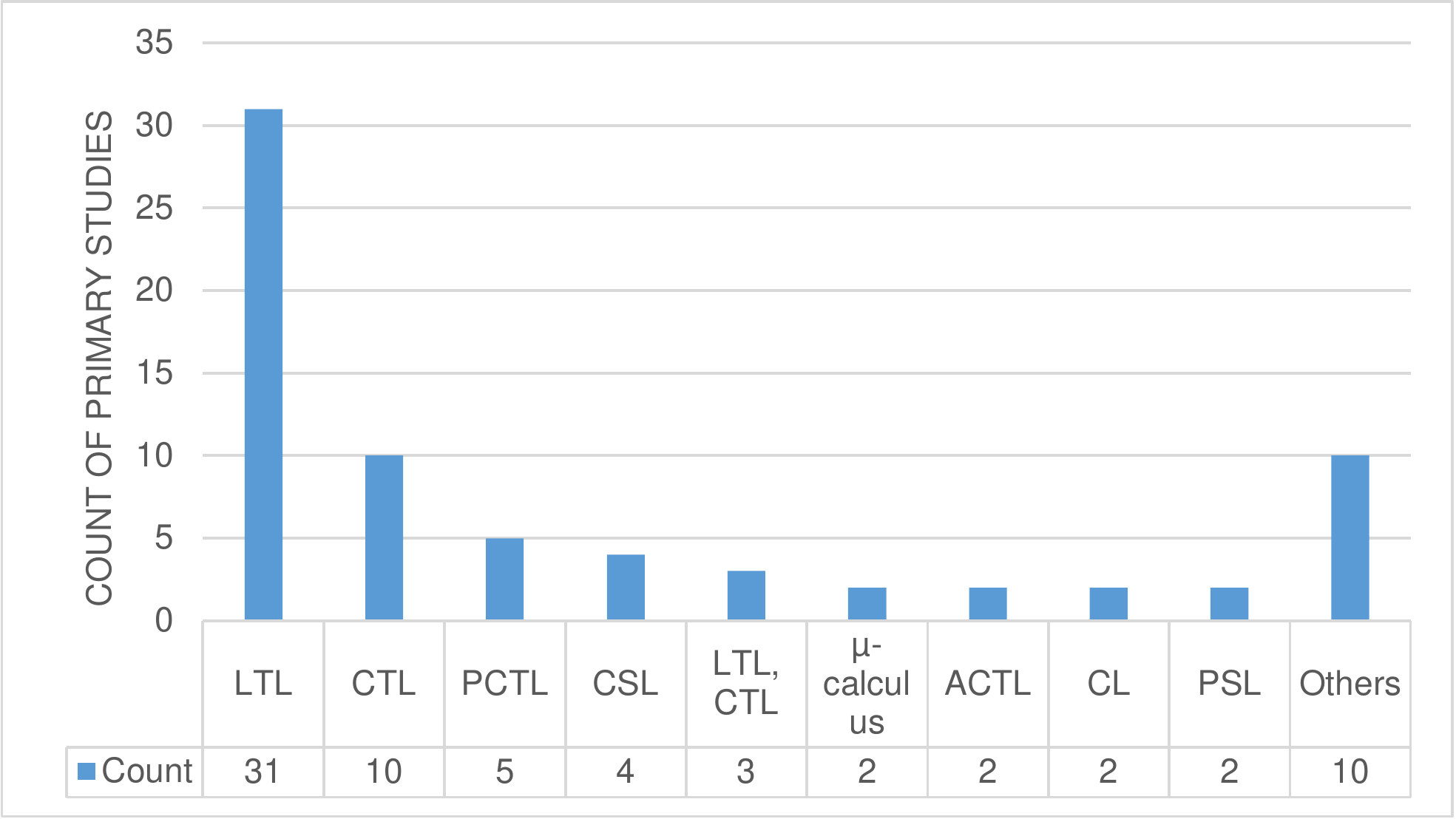}
	\subcaption{Temporal logics used for specifications. Formalisms that only occur once in the surveyed literature are combined into the category \enquote{Others} (FLTL, TCTL, Datalog, FOL, ALCCLTL, XCTL, CTLK, ATL, and STLCS).}
	\label{fig:spec:formalism}
\end{subfigure}
~
\begin{subfigure}[t]{0.3\linewidth} 
	\includegraphics[scale=.4]{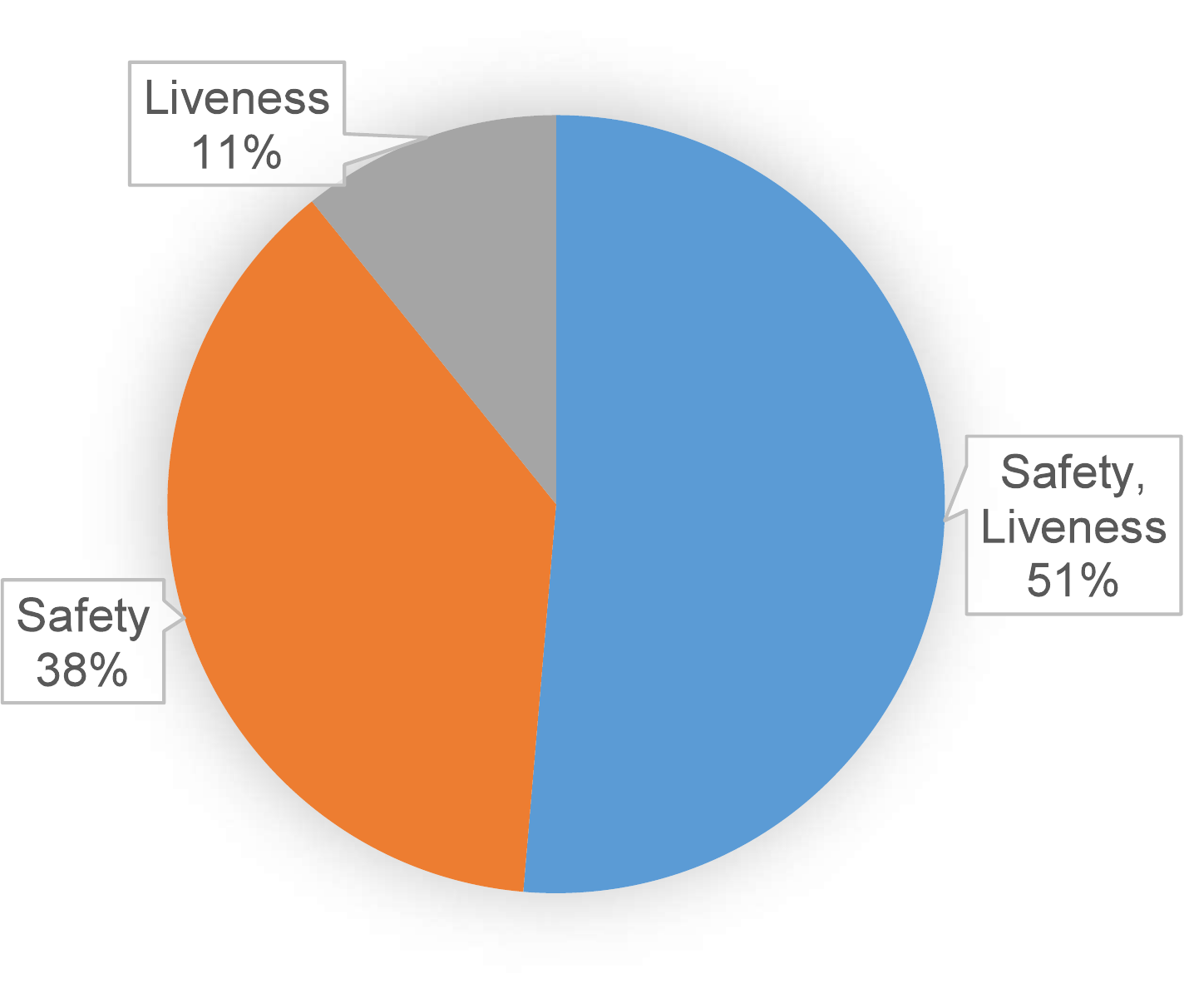}
	\subcaption{Types of specification properties. Safety properties are considered in almost 90\% of all primary studies.}
	\label{fig:spec:properties}
\end{subfigure}
\caption{\rsq{5} Types of system specifications and specification properties.}
\end{figure}

\noindent
\fig{spec:formalism} shows the results for the temporal logics used to specify requirements in the primary studies.\footnote{The primary studies for each temporal logic are listed in \tabrqfivea.}
\specltl primary studies (27\% of all primary studies) use \emph{\gls{LTL}} and \specctl (9\%) use \emph{\gls{CTL}}.
For both cases, approaches that use graphical representations of counterexamples exist (\eg \cite{KanavA17,AungNO18a,Nguyen017-1,PhyoO18} for \gls{LTL}, and \cite{PatilVP15,JinRS04,SchinzTMW04} for \gls{CTL}). The frameworks NuSeen~\cite{ArcainiGR17a}, PLCverif~\cite{DarvasVA15}, and MODCHK~\cite{PakonenMLK13} support both \gls{LTL} and \gls{CTL}.
Furthermore, \specpctl primary studies (4\% of all primary studies) use \emph{\gls{PCTL}} and \speccsl (4\%) use \emph{\gls{CSL}} to tackle probabilistic specifications. The \specpctl studies~\cite{JansenAVWKB12,AljazzarL10,JansenAKWKB11,AbrahamJWKB10,DebbiB13-1} that use \gls{PCTL} minimize counterexamples while the \specpctl studies~\cite{KuntzLL11,AljazzarLLS11,Leitner-FischerL13-2,Leitner-FischerL13-3}  that uses \gls{CSL} generate fault trees. 

Other specification languages we found in our survey (two primary studies for each) are \emph{\gls{CL}}, \emph{ACTL}, \emph{\gls{PSL}}, and \emph{$\mu$-calculus}. \gls{CL} is inspired by dynamic, temporal, and deontic logic and is used for specifying contracts containing clauses. Angeloc \etal~\cite{AngelovCS13} convert the \gls{CNL} language into \gls{CL} using the \gls{GF} and perform verification with the \gls{CLAN} model checker. The errors found in the counterexample are then highlighted in the \gls{CNL}. Frameworks like AutoFocus~3~\cite{CampetelliHN11} and PyNuSMV~\cite{BusardP18} support the $\mu$-calculus along with \gls{LTL}, \gls{CTL}, CTLK and ATL.
ACTL and \gls{PSL} logics are verified using the \gls{NuSMV} model checker. Clarke \etal~\cite{ClarkeJLV02} use ACTL to generate multiple counterexamples and Shen \etal~\cite{ShenQL05-3} uses ACTL for counterexample minimization.

\subsubsection{Types of specification properties}
\label{sec:analysis:spec:properties}

\noindent
Specifications may be used to specify \emph{safety} and \emph{liveness} properties. Safety properties assert that something bad never happens, while liveness properties assert that something good will eventually happen~\cite{Sistla94}. A counterexample to a liveness property in a finite system is lasso-shaped, which consists of a prefix that leads to a loop. For safety properties, a counterexample is finite in length~\cite{BiereAS02}. As shown in \fig{spec:properties}, most primary studies address safety properties (38\%), or safety and liveness properties (51\%) at the same time~\cite{EdelkampLL01,EdelkampLL04,SchuppanB05} while only 11\% address the liveness properties alone.\footnote{The primary studies for each type of property are listed in \tabrqfiveb.} \change{Out of the 51\% of studies that address safety \emph{and} liveness properties, the majority of studies (32\%) represent a counterexample in a graphical format, \eg animate the counterexample with a loop~\cite{PakonenBV18,BeerBCOT12,NguyenO17}. The remaining 19\% follow different methods to process or represent the counterexample generated for \emph{safety} and \emph{liveness} properties, \eg Edelkamp \etal~\cite{EdelkampLL04} use a nested depth-first search algorithm for analyzing liveness properties, but directed search for safety properties to generate a minimized counterexample.}

Mostly the specification of liveness properties makes use of inevitable executions. Therefore, inevitable execution properties belong to a class of liveness properties~\cite{Barbon0S18}. The work by Barbon \etal~\cite{Barbon0S18} focus on improving the comprehension of counterexample specifically for such inevitability properties, which is predominantly used by developers in practice~\cite{DwyerAC99}.

\subsubsection{Answer to RQ5}
\label{sec:analysis:spec:conclusion}

\noindent
In our survey, we found that 44 (38\% of the primary studies) use either LTL, CTL, or both, and 10 (9\%) use probabilistic specifications. However, just one primary study deals with counterexample explanation of real-time specifications~\cite{GerkingSDH15}, in this case based on UPPAAL~\cite{BengtssonLLPY96,LarsenPY97}.
Finally, 51\% of the primary studies verify both safety and liveness properties while 38\% focus exclusively on safety properties and 11\% on liveness properties. 

%% file: tex/rsq_tools.tex
\subsection{\rsq{6}Which verification tools and frameworks are developed and used to explain counterexamples, and how do they effect the counterexample explanation?}
\label{sec:analysis:tools}

\noindent
To answer this research question, we collected the verification tools and frameworks used to develop counterexample explanation approaches.
These results are of particular interest for practitioners looking for reuse of existing solutions. Furthermore, we collected qualitative data to demonstrate the effect of tools and frameworks on the counterexample explanation.
%
\change{Luteberget and Johansen~\cite{LutebergetJ18} state that \blockquote{\textit{a verification tool which runs invisibly alongside the design, giving feedback on the current state of the design at any time could have a higher impact on the design process}}.}

\subsubsection{Verification tools}
\label{sec:analysis:tools:mc}

\begin{figure}
	\begin{subfigure}[b]{\linewidth}  
		\includegraphics[width=\textwidth]{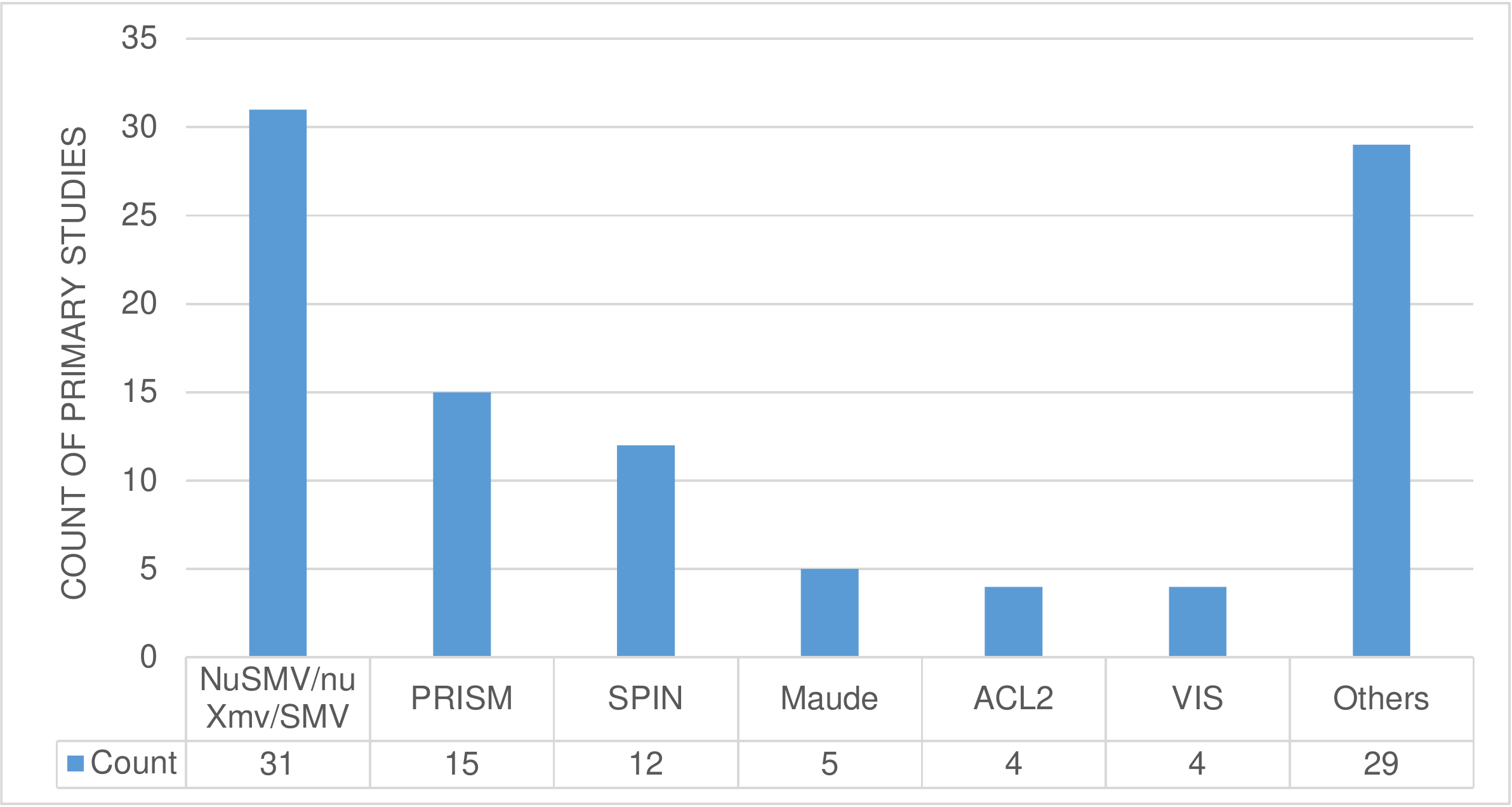}
		\subcaption{Verification tools used by the primary studies. Tools that with less than four occurrences are clustered in the category \enquote{Others} (Zchaff, Z3, CADP, ViVe/SESA, topochecker, SpinJa, SLDV, SLAM, Yices, MTSA, LTSA, CPAchecker, Ultimate Automizer, CLAN, CBMC, JPF, XCheck, SAL, CoCoSim, JKind, MCMAS+, and MiniSAT).}
		\label{fig:tools}
	\end{subfigure}
	\hfill
	\begin{subfigure}[b]{\linewidth}  
		\includegraphics[width=\textwidth]{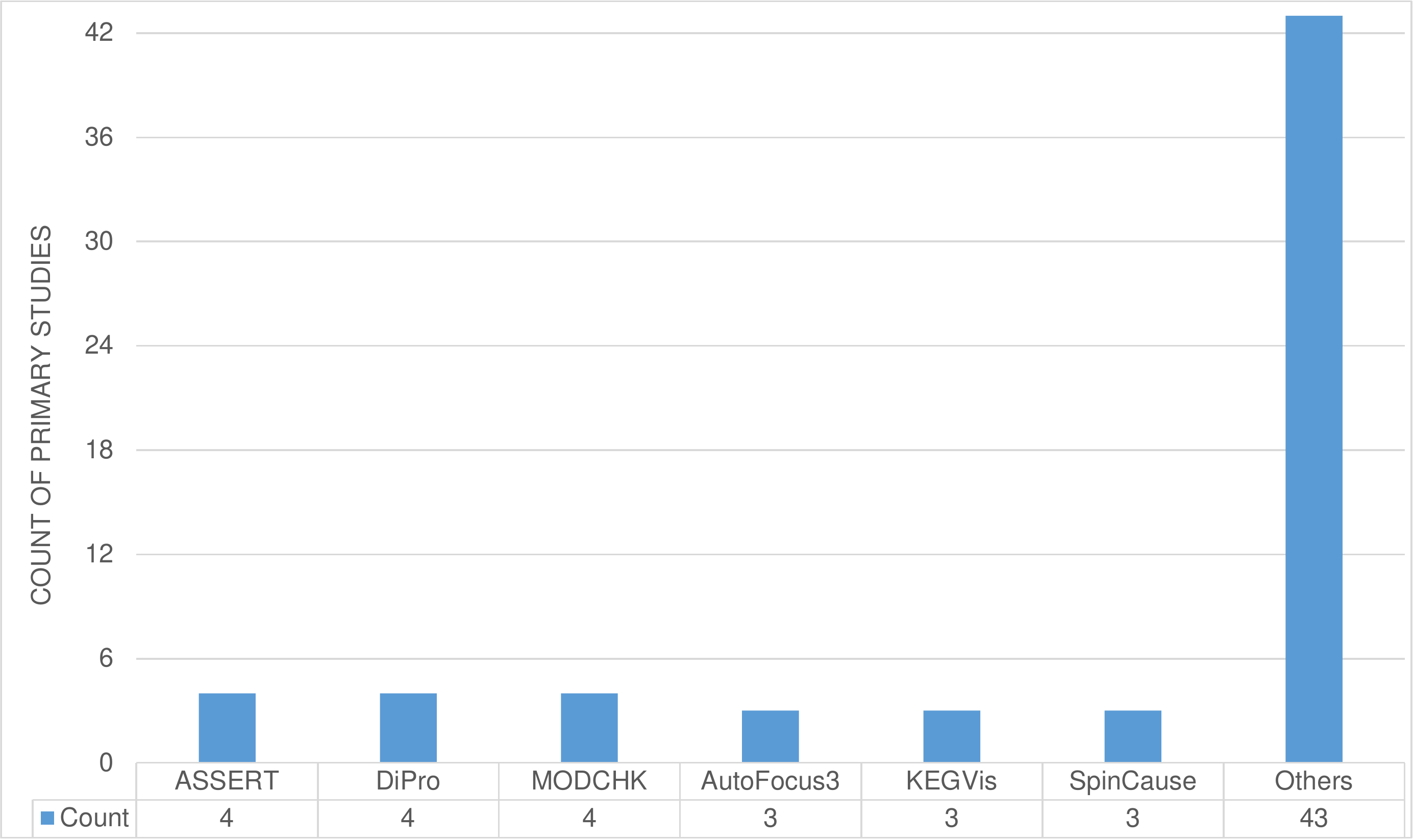}
		\subcaption{Frameworks for counterexample interpretation. Frameworks that occur less than three primary studies are combined into the category \enquote{Others} (IFADIS, STANCE, FAULTCAT, CX2FT, SpinRCP, MechatronicUML, AnaCon, RailComplete, [Mc]SQUARE, Pseudo-merge, EOFM, ProofProd, Evidence Explorer, ELARVA, Arcade.PLC, SMART, Alfi, PyNuSMV, RuleBase PE, COMICS, AMASE, NuSeen, FASTEN, PLCverif, Ivy, DSValidator, IBM RoseRT, Theseus, VIS, MACEMC, FLAVERS/Ada, OERITTE, and GraphML).}
		\label{fig:frameworks}
	\end{subfigure}
\caption{\rsq{6} Frameworks and verification tools used by primary studies.}
\end{figure}

\noindent
\fig{tools} shows the verification tools used by the primary studies.\footnote{The primary studies for each verification tool are listed in \tabrqsixa.}
The predominantly used model checker in the context of counterexample explanation is the \emph{\gls{NuSMV}/nuXmv/\gls{SMV}}. It is used by \toolsnusmv primary studies (27\% of all primary studies). 
nuXmv extends \gls{NuSMV} that extends \gls{SMV}~\cite{Mcmillan93} and they can verify specifications expressed in \gls{LTL}, \gls{CTL}, \gls{PSL}, or \gls{RTCTL}. 
We found two particular primary studies that use \gls{NuSMV}.
Schuppan and Biere~\cite{SchuppanB05} identify the shortest lasso-shaped counterexample for \gls{LTL}, and Pakonen \etal~\cite{PakonenBV18} highlight the value of atomic proposition that are important for understanding counterexample. 

\toolsspin primary studies use \emph{\gls{SPIN}}~\cite{Holzmann97,Holzmann04} (10\% of all primary studies) and \toolsmaude use \emph{Maude}~\cite{EkerMS02,EkerMS03,ClavelDELMMT07} (4\%) to verify \gls{LTL} specifications.
For instance, \gls{SPIN} 
is used with minimizing counterexamples~\cite{EdelkampLL01,GastinMZ04,GastinM07,HansenG08}.
Maude integrated with Draw-SVG is used to animate user-defined state machines%
~\cite{Nguyen017,AungNO18a,Nguyen017-1,PhyoO18}.
Furthermore, \emph{\gls{VIS}}~\cite{BraytonHSSACEKKPQRSSSV96,JeongYC10} is used by \toolsvis primary studies (3\% of all primary studies) and \gls{ACL2} by \toolsvis (3\%).
\gls{VIS} supports verification of fair \gls{CTL} specifications, simulation of logic circuits, and explanation of counterexample in a flow chart representation.
\emph{\gls{ACL2}}~\cite{ChamarthiDMV11,KaufmannMPM13} is a tool for modeling, simulation, and theorem proving. 
The \gls{ASSERT} tool integrates \gls{ACL2} with ontologies~\cite{CrapoM19,CrapoMMR17,MoitraSCCDLYMM18,MoitraSCDLMMM19} to analyze an incomplete set of specifications and localize the error by identifying which specifications are responsible for the error.

Model checkers like \emph{\gls{PRISM}}~\cite{HintonKNP06,KwiatkowskaNP11} and \emph{\gls{MRMC}}~\cite{KatoenKZ05,KatoenZHHJ11} are used to verify probabilistic specifications. \toolsprism primary studies (13\% of all primary studies) use \gls{PRISM} and among them \toolsmrmc primary studies~\cite{AljazzarLLS11,WimmerJAKB14,AbrahamJWKB10} use both \gls{MRMC} and \gls{PRISM} for probabilistic model checking, and particularly with minimization of counterexamples~\cite{AljazzarLLS11,AbrahamJWKB10}. Leitner-Fischer and Leue~\cite{Leitner-FischerL13-3}, and Kuntz \etal~\cite{KuntzLL11} use the FaultCAT and CX2FT frameworks to generate fault trees with the help of counterexample generated by \gls{PRISM}. Similarly, Leitner-Fischer and Leue~\cite{Leitner-FischerL13-2,Leitner-FischerL14} use the SpinCause framework to generate fault trees with \gls{SPIN} and \gls{PRISM}. 

\subsubsection{Frameworks}
\label{sec:analysis:tools:frameworks}

\noindent
In our survey, we did not find any framework that is reused by more than four primary studies (\fig{frameworks}).\footnote{The frameworks and the primary studies that use them are listed in \tabrqsixb.}
The \emph{\gls{ASSERT}} framework used by \fwassert primary studies (3\% of all primary studies) check the requirements for conflicts and completeness by using an automated theorem prover~\cite{CrapoM19,CrapoMMR17,MoitraSCCDLYMM18,MoitraSCDLMMM19}. Further, it generates requirements-based test cases using \gls{SMT}. Moitra \etal~\cite{MoitraSCCDLYMM18} state that \blockquote{\gls{ASSERT} saves time and cost by identifying errors early in the development process and automating requirements-based test generation}.
 
Out of all frameworks shown in \fig{frameworks}, only \emph{\gls{DiPro}} that is used by 4 primary studies (3\% of all primary studies) and \emph{SpinCause} that is used by 3 studies (3\%) support verification of probabilistic systems. The open-source tool DiPro is used with \gls{PRISM} and \gls{MRMC} for the computation and graphical representation of probabilistic counterexamples for \glspl{DTMC}, \glspl{CTMC} and \glspl{MDP}~\cite{AljazzarLLS11}. SpinCause is based on SpinJa~\cite{JongeR10}, a Java re-implementation of 
\gls{SPIN}~\cite{Leitner-FischerL13-2}. Florian and Leue~\cite{Leitner-FischerL14} use SpinCause for causality checking of \gls{PROMELA} and \gls{PRISM} models and the approach is evaluated with industry sized models.
 
Two frameworks that support animating counterexamples are \emph{MODCHK}~\cite{PakonenTHP17,PakonenB17,PakonenBV18} and \emph{AutoFocus~3}~\cite{CampetelliHN11,CampetelliJBDKZ15,KanavA17}. 
\fwmodchk primary studies (3\% of all primary studies) use \emph{MODCHK} to animate a counterexample in the user-provided input function block diagram. Similarly, \emph{AutoFocus~3} used by \fwautofocus primary studies (3\%) animate a counterexample in the user-provided component diagrams. 

\emph{\gls{KEGVis}} used by \fwkegvis primary studies (3\% of all primary studies) presents witnesses graphically with the daVinci Presenter~\cite{FrohlichW94} for layout and exploration~\cite{GurfinkelCD03}. Besides simply browsing the witnesses, the user can customize the exploration strategy, for instance, in terms of forward and backward exploration, adjusting step granularity, and choosing witnesses based on size.
Particularly, \gls{KEGVis} is used together with the model checkers \gls{XChek}
and \gls{NuSMV} in~\cite{ChechikG07,GurfinkelC03,GurfinkelCD03}.

In the following, we summarize notable frameworks that are used by less than three primary studies.

Gerking \etal~\cite{GerkingSDH15} present a model-to-model transformation from a design model (in-terms of domain-specific model checking~\cite{VisserDW12}) to a verification model that is based on the input language of the model checker UPPAAL. The key feature of this approach is to bridge large differences between the domain-specific modeling language and the model checker's input language. Thus, it also translates counterexamples back to the level of the domain-specific models. A similar approach is followed for the AutoFocus~3~\cite{KanavA17,CampetelliJBDKZ15} and \gls{FASTEN}~\cite{RatiuGS19} frameworks.

\emph{SMGA} is an animation tool that visualizes counterexamples generated by Maude as a \emph{state machine}~\cite{AungNO18a}. 
In the same way, Nguyen~\cite{Nguyen017}, Phyo and Ogata~\cite{PhyoO18}, as well as Nguyen and Ogata~\cite{NguyenO17,Nguyen017-1} allow the user to design picture of animations, adjust the speed of animations, and select states that satisfy some given conditions and/or constraints from a finite computation. 
\emph{Theseus} provides two animation options~\cite{GoldsbyCKK06}: automatic playback and an incremental playback option, by using \emph{state diagram} animation and \emph{sequence diagram} generation respectively. The counterexample is animated by automatic playback while the incremental playback animates the counterexample in a step-by-step manner. Similarly, \emph{SpinRCP}~\cite{BrezocnikVV14} transforms a Spin simulation trail into a standard \gls{MSC} and supports interactive simulations. 

\emph{\gls{STANCE}} is a framework that is integrated with the Simulink toolset that visually shows the structural parts of Simulink models. The tool thus filters out the irrelevant data from the counterexample, thus easing the job of the engineers to focus only on interesting details of its execution~\cite{BochotVWW10,CastillosWW15}. Similarly, the \emph{CLEAR}~\cite{BarbonLS19} tool highlights actions that causes violation in the counterexample and specifically supports the debugging of concurrent systems.

Gerking \etal~\cite{GerkingSDH15} present a model-to-model transformation from a design model (in-terms of domain-specific model checking~\cite{VisserDW12}) to a verification model that is based on the input language of the model checker UPPAAL.

\subsubsection{Answer to RQ6}
\label{sec:analysis:tools:conclusion}

\noindent
Our results show that the use of verification tools and specification languages (\sect{analysis:spec}) are clearly coupled, because certain languages can only be verified by certain tools. Since \gls{LTL} is the most used logic (38\%), the verification tools \gls{NuSMV}/nuXmv/\gls{SMV}, \gls{SPIN} and Maude are found to be used by 41\% of the primary studies (all of them can verify \gls{LTL} specifications). 13\% of primary studies use probabilistic model checkers such as PRISM and MRMC.

Each of the most used frameworks \gls{ASSERT}, \gls{DiPro} or MODCHK is only used by \toolsreuse primary studies. 
Frameworks such as MODCHK, AutoFocus~3, SMGA, FASTEN, and Theseus support animating the counterexample in a user-given design model. By performing the animation, engineers can understand the counterexample by executing it stepwise. Particularly, AutoFocus~3 and FASTEN as well as the approach by Gerking \etal~\cite{GerkingSDH15} support minimizing the counterexample along with explaining the counterexample in a graphical format.

%% file: tex/rsq_evaluation.tex
\subsection{\rsq{7} How are counterexample explanation approaches evaluated?}
\label{sec:analysis:evaluation}

\noindent
Among all primary studies, 97 studies (84\%) perform an evaluation. Among these 97 studies, we have identified the application domain in 69 studies (71\%), while 28 studies (29\%) do not target a specific domain. Similarly, 87 of the 97 studies (90\%) mention the types of applications used for the evaluation.

\subsubsection{Application domains}
\label{sec:analysis:evaluation:domains}

\noindent
\fig{evaluationd} shows the results for the application domains used for evaluating counterexample explanation approaches by the primary studies.\footnote{The primary studies for each application domain are listed in \tabrqsevena.}
\evaldomaincomm primary studies (14\% of all studies that perform an evaluation) perform the evaluation on \emph{communication protocols}, \eg to evaluate graphical counterexample representations~\cite{NguyenO17,PadonMPSS16,AungNO18a,Nguyen017-1,PhyoO18} or counterexample minimization~\cite{EdelkampLL04,HansenK06,ZhaoJC11}. 
\evaldomaincomm~further primary studies (14\%) perform the evaluation in the \emph{hardware} domain, \eg with applications such as \gls{PLC} software~\cite{PakonenMLK13,DarvasVA15,BiallasFSK15} and circuits~\cite{CoptyIWKK03,ShenQL05-2}.

In total 28 primary studies (29\%) evaluate their approach in a safety-critical application domain such as \emph{automotive}~(\evaldomainautomotive), \emph{robotics}~(\evaldomainrobotics), \emph{avionics}~(\evaldomainavionics), \emph{nuclear}~(\evaldomainnuclear), and \emph{railway}~(\evaldomainrailway). Evaluation of textual representations is performed, \eg in avionics~\cite{CrapoM19,CrapoMMR17,MoitraSCCDLYMM18,MoitraSCDLMMM19}, the railway domain~\cite{LutebergetCJS17}, and robotics~\cite{FengGCT18}. Evaluation of graphical representations is performed, \eg in the automotive~\cite{Leitner-FischerL14,CastillosWW15}, and nuclear domain~\cite{PakonenBV18,PakonenTHP17,PakonenB17,PakonenBB21}.

\subsubsection{Types of Applications}
\label{sec:analysis:evaluation:usecases}

\noindent
We categorize the applications used for the evaluation in primary studies into three groups listed below.
We additionally distinguish between primary studies that introduce novel applications and those that reuse existing applications.

\begin{enumerate}\itemsep0em
	\item \textit{Industrial application:} The evaluation uses a real-world industrial application.
	\item \textit{Non-industrial application:} The evaluation uses a real-world non-industrial application such as an application developed in a research lab.
	\item \textit{Example application:} The evaluation uses a simple and small exemplary application such as a toy example.
\end{enumerate}

\noindent
\fig{evaluationc} shows the results for these types of applications.\footnote{The primary studies for each application type are listed in \tabrqsevenb.}
\evalusecaseind primary studies (16\% of all studies that perform an evaluation) are evaluated with \emph{industrial applications}, which can be considered a good indication of their maturity and scalability. Industrial applications range from Finnish nuclear industry~\cite{PakonenTHP17} and aero-engine blade forging~\cite{ZhengTZ16} to desalination plants~\cite{CampetelliJBDKZ15}.
Approaches are also evaluated by \textit{referring to existing industrial applications} in \evalusecaseindref primary studies (23\%). Examples are the airbag system by Aljazzar \etal~\cite{AljazzarFGKLL09} and referred by Kuntz \etal~\cite{KuntzLL11} and Aljazzar \etal~\cite{AljazzarLLS11}, and a flasher manager application by Collavizza \etal~\cite{CollavizzaVPRR14} and referred by Castillos \etal~\cite{CastillosWW15}.

Likewise, \evalusecasenonindref primary studies (32\%) evaluate their approaches \emph{referring to existing non-industrial applications}, \eg to evaluate approaches to counterexample minimization~\cite{SchuppanB05,EdelkampLL04,HansenG08,TanACZL04,AbrahamJWKB10,DebbiB13-1,ZhaoJC11}, graphical representations~\cite{PadonMPSS16,ElamkulamGRKGKDM06,AungNO18a,Nguyen017-1,PhyoO18}, and textual representations~\cite{FengGCT18,LutebergetCJS17,GroceCKS06,AngelovCS13}. An evaluation based on \emph{non-industrial applications} is performed by \evalusecasenonind primary studies (7\%), \eg to evaluate some of the textual representation approaches~\cite{FengGCT18,LutebergetCJS17,GroceCKS06,AngelovCS13}.

Finally, \evalusecaseexample primary studies (18\%) use \emph{example applications}. Mostly, such applications are used to demonstrate the proposed approach, \eg for graphical representations of counterexamples~\cite{ElamkulamGRKGKDM06,Nguyen017,OvsiannikovaBPV20}.

\subsubsection{Evaluation aspects}
\label{sec:analysis:evaluation:aspects}

\noindent
We distinguish three types of aspects that are the goals of the evaluation~\cite{WohlinRHOR00}:

\begin{enumerate}\itemsep0em
	\item \textit{Efficiency and Performance:} Evaluation focuses on the efficiency and computational demand such as execution time of approaches.
	\item  \textit{Effectiveness:} Evaluation focuses on effectiveness and usability of counterexample explanation approaches.
	\item \textit{Scalability:} Evaluation investigates whether an approach is scalable and can be applied to complex scenarios and systems.
\end{enumerate}

\begin{sidewaysfigure}
	\begin{subfigure}{0.5\linewidth}
	\centering
	\includegraphics[width=\linewidth]{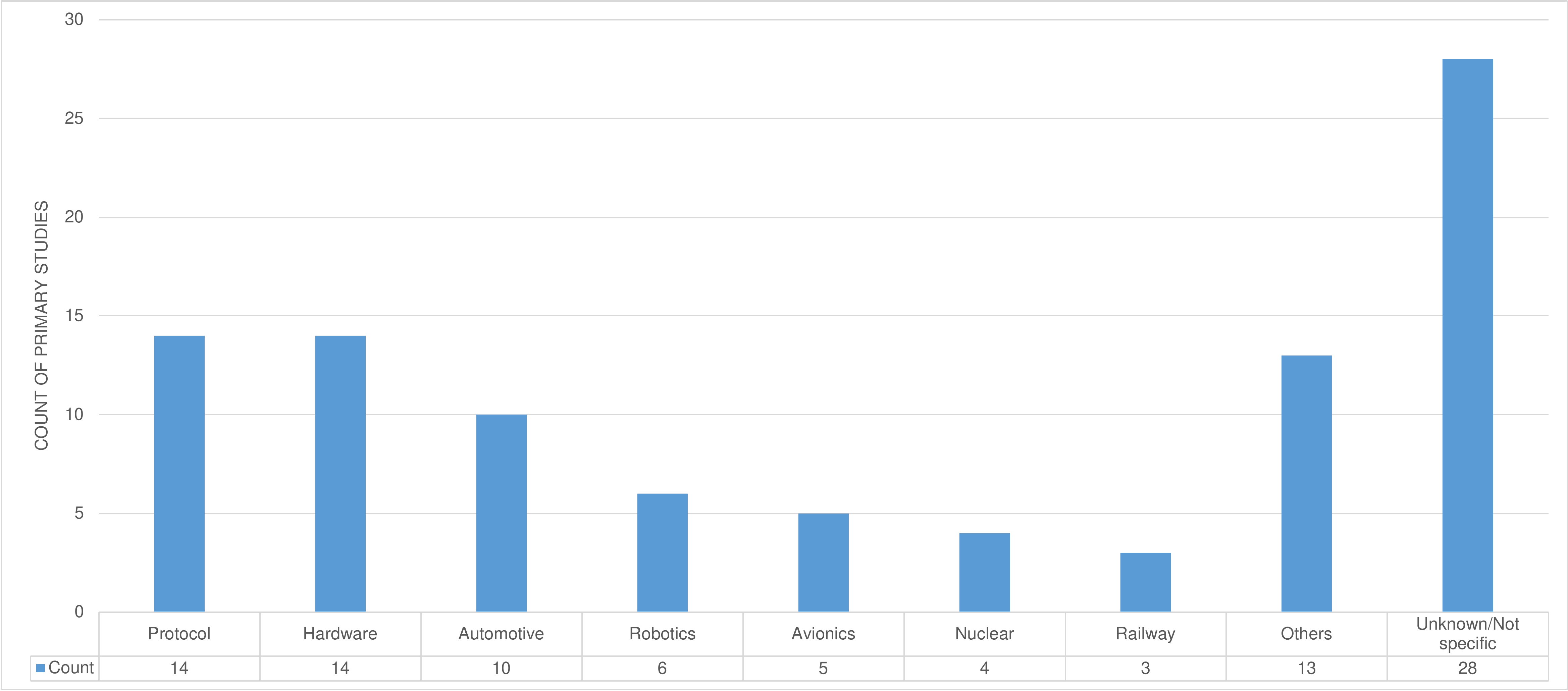}
	\subcaption{Overview of the different application domains. Domains that were found in less than three primary studies, namely, Internet Service, Desalination Plant, Library, Billing Renewal, GPS Data, and Operating System are grouped as "Others".}
	\label{fig:evaluationd}
\end{subfigure}
~
\begin{subfigure}{0.5\linewidth}
	\centering
	\includegraphics[width=\linewidth]{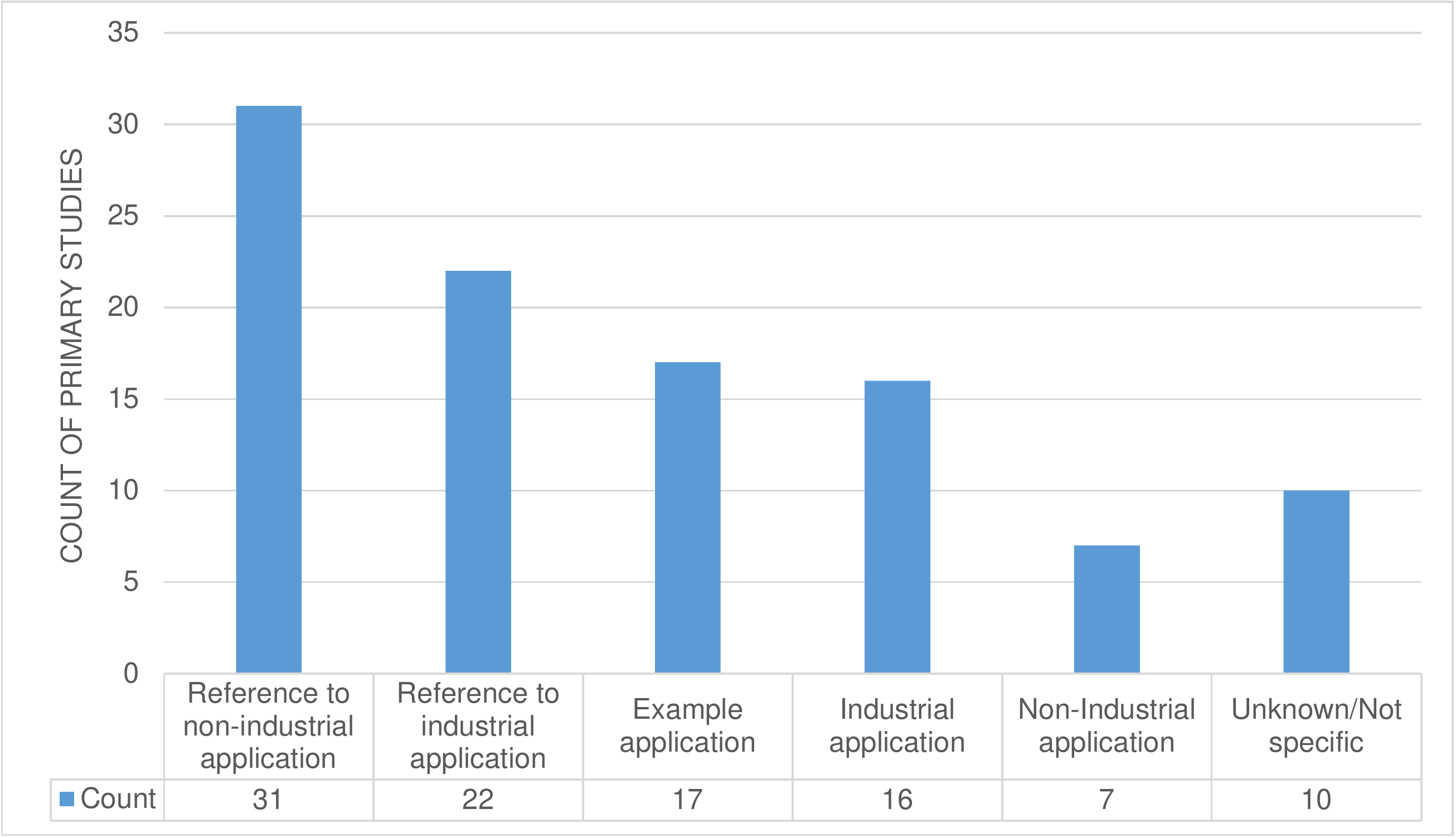}
	\caption{Types of applications.}
	\label{fig:evaluationc}
\end{subfigure}

\begin{subfigure}{0.5\linewidth}
	\centering
	\includegraphics[width=\linewidth]{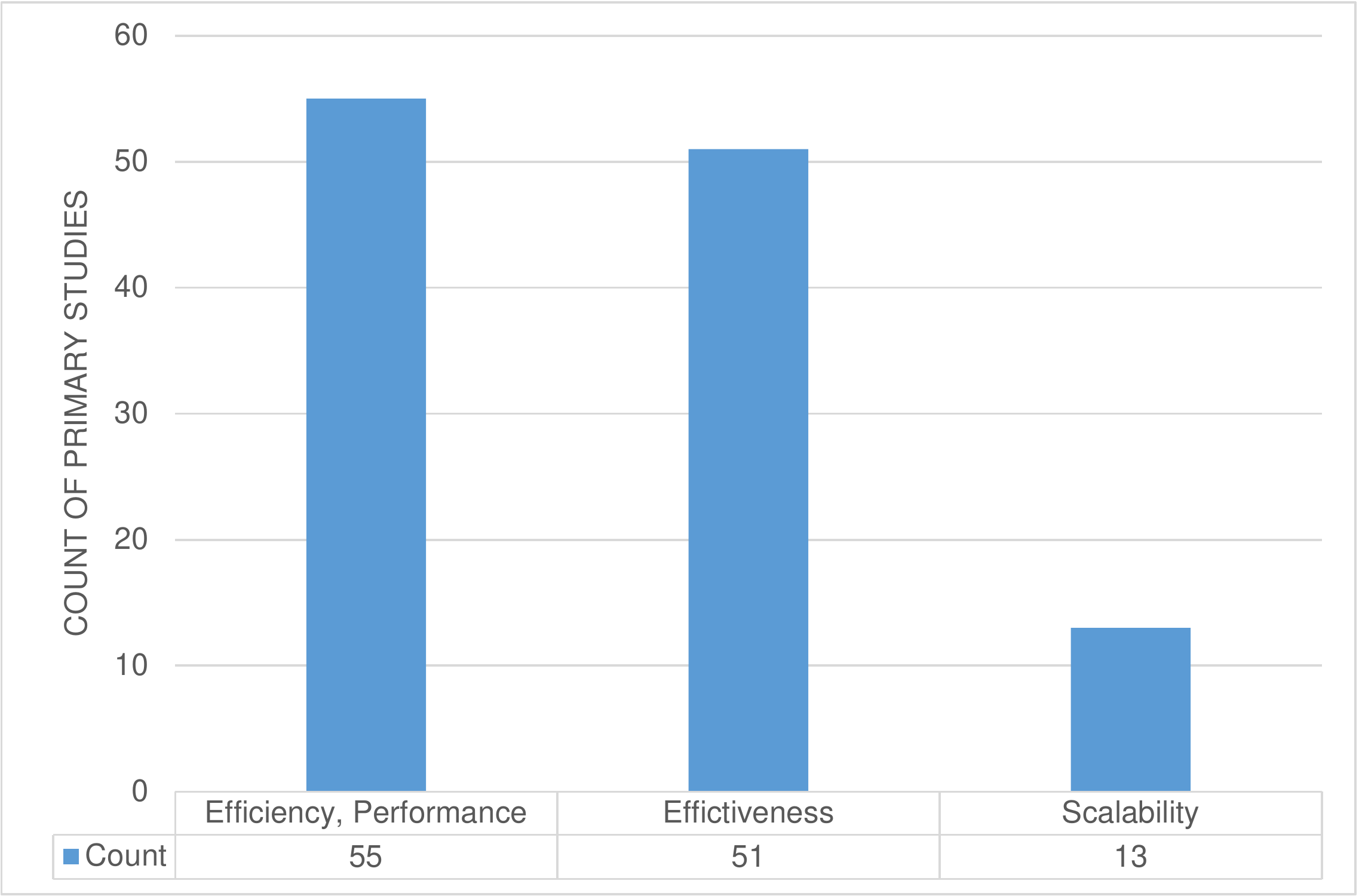}
	\caption{Overview of evaluation aspects.}
	\label{fig:evaluationa}
\end{subfigure}
~
\begin{subfigure}{0.5\linewidth}
	\centering
	\includegraphics[width=\linewidth]{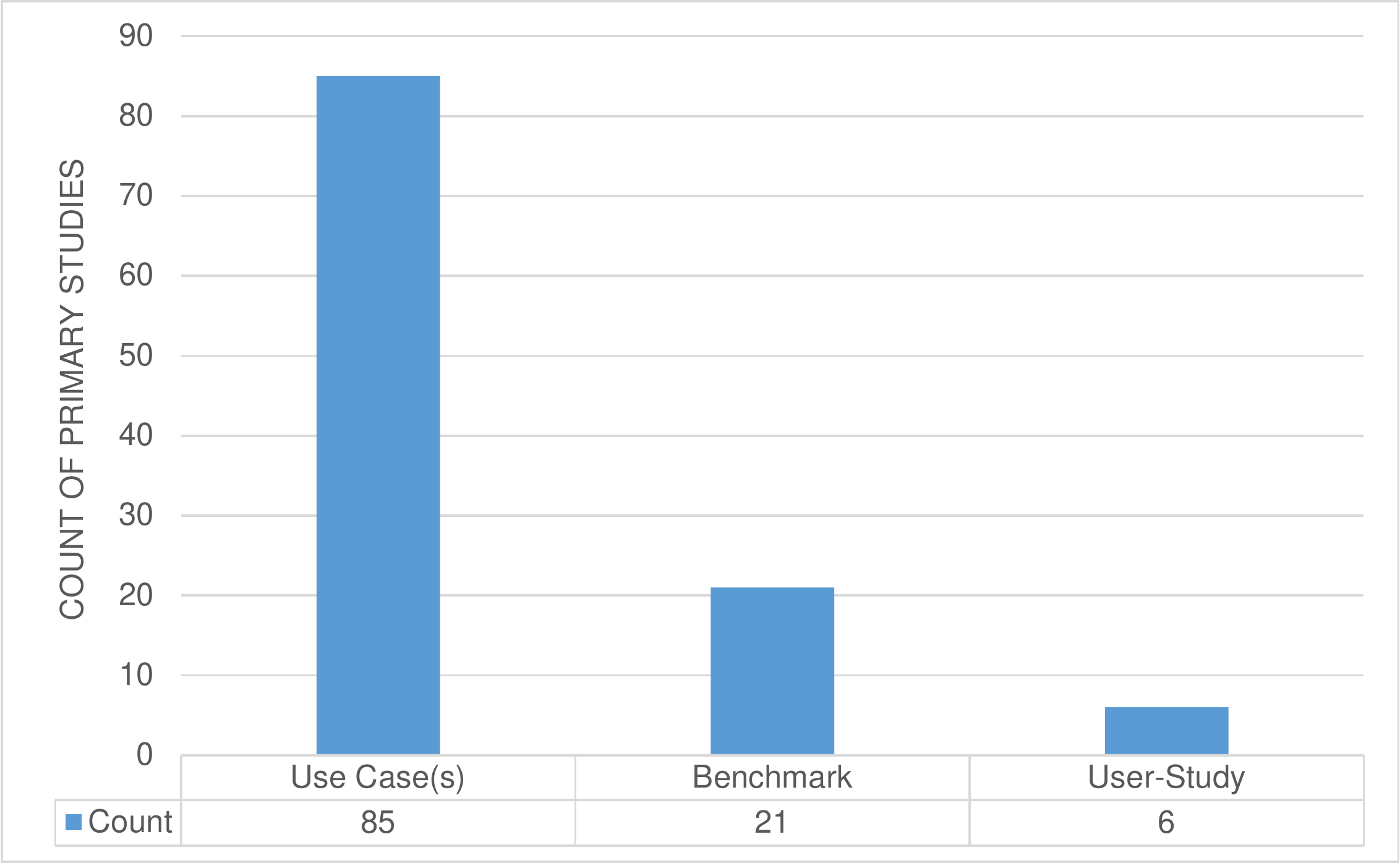}
	\caption{Overview of evaluation methods. }
	\label{fig:evaluationb}
\end{subfigure}
\caption{\rsq{7} Results for the application domain, type of application, evaluation aspect, and evaluation method. For each of them, if a primary study matches more than one category (\eg  it uses several evaluation methods), the study is counted for each category (\ie for each evaluation method in this example).}
\end{sidewaysfigure}

\noindent
\fig{evaluationa} shows the distribution of the evaluation aspects among the primary studies.\footnote{The primary studies for each evaluation aspect are listed in \tabrqsevenc.}
The aspect of \emph{efficiency/performance} is evaluated in \evalaspectperform primary studies (57\% of all studies that perform an evaluation). 
Two exemplary studies are the ones by Leitner-Fischer and Leue~\cite{Leitner-FischerL14} and Tan \etal~\cite{TanACZL04}.
The run-time of the proposed approaches is noted with different use cases for minimizing counterexample using the DFS and BFS search methods.

Close to the count of efficiency/performance, \emph{effectiveness} is evaluated in \evalaspecteffect primary studies (53\%). The effectiveness of frameworks IFADIS~\cite{LoerH06} and FASTEN~\cite{RatiuGS19} are evaluated in user studies. 
In our survey we also found \evalaspectperformeffect additional primary studies that evaluate approaches focusing on both \emph{efficiency/performance and effectiveness}~\cite{PakonenBV18,LutebergetCJS17,PatilVP15,LeueB12,BeyerDDHS15}.

An evaluation focusing on \emph{scalability} is performed by \evalaspectscale primary studies (13\%). Two examples are the studies by Dominguez \etal~\cite{DominguezD13} focusing on generating multiple counterexamples, and by Barbon \etal~\cite{BarbonLS18} focusing on minimizing counterexamples for concurrent systems. The proposed approaches are evaluated with several use cases that are different in size and complexity to show the scalability of the approach.

\subsubsection{Evaluation methods}
\label{sec:analysis:evaluation:methods}

\noindent
We distinguish three different types of evaluation methods~\cite{WohlinRHOR00}:

\begin{enumerate}\itemsep0em
	\item \textit{Use Case:} Anecdotal evidence, the evaluation aspect is applied to one or more use cases and the results are reported.
	\item \textit{Benchmark:} An evaluation aspect is compared to a particular baseline or any established approach.
	\item \textit{User Study:} Evaluation from the user's point of view, where feedback is collected from users, \eg domain experts.
\end{enumerate}

\fig{evaluationb} shows the results for the evaluation methods used by the primary studies.\footnote{The primary studies for each evaluation method are listed in \tabrqsevend.}
Anecdotal evidence based on \emph{use cases} is the most used method, found in \evalmethodusecase primary studies (88\% of all studies that perform an evaluation), \eg 
to evaluate graphical~\cite{JeongYC10,GoldsbyCKK06,PakonenB17,SchinzTMW04,CampetelliJBDKZ15} or textual representations~\cite{LutebergetJ18,CrapoMMR17}.

An evaluation with a \emph{benchmark} is performed by \evalmethodbenchm primary studies (22\%), mostly to evaluate counterexample minimization~\cite{GastinMZ04,JansenAVWKB12,RaviS04,EdelkampLL04,HansenG08,HansenK06}.
Notably, Beyer \etal~\cite{BeyerDDHS15} created benchmarks for 3964 verification tasks from all categories of SV-COMP 2015~\cite{Beyer15}. Chang \etal~\cite{ChangBM07} evaluate their trace minimization approach with nine benchmark designs.

However, only \evalmethoduserstudy primary studies (6\%) perform a \emph{user study} to assess the effectiveness of counterexample explanation approaches.
Barbon \etal~\cite{BarbonLS18} perform a user study with 17 developers to assess their approach to counterexample minimization. The developers are asked to analyze the original and the minimized counterexample. The results of this use study shows that developers spend more time on analyzing the original counterexample than the minimized counterexample.
Moitra \etal~\cite{MoitraSCCDLYMM18} state that \enquote{Having engineers from the business units embedded into the team developing the tools was important because it kept us focused on solving the real problems the business was having.}

\subsubsection{Answer to RQ7}
\label{sec:analysis:evaluation:conclusion}

\noindent
A particular striking point is that almost 29\% of the primary studies evaluate their approaches in a safety-critical domain. Furthermore, the number of approaches evaluated with industrial applications is quite significant (39\% of all studies that perform an evaluation). This underlines the applicability of the evaluated approaches in critical real-world applications. Efficiency and performance as the evaluation aspect is addressed by 57\% while effectiveness is evaluated by 53\% of the primary studies. The most used evaluation method is the \textit{use case}, occurring in 88\% of the primary studies.

%% file: tex/ThreatsToValidity.tex
\section{Threats To Validity}
\label{sec:ttv}

\noindent
There are several potential validity threats to the design of our systematic literature review. We discuss the external validity of our study and reliability of our analysis according to Wohlin \etal~\cite{WohlinRHOR00}.

\subsection{External Validity}
\label{sec:ttv:external}

\noindent
To mitigate threats to the external validity of the survey, we define a systematic selection process as documented in \sect{process} and depicted in \fig{process}.
The process resorts to the huge literature database \emph{Google Scholar}.
Google Scholar weights query results by the number of citations, which introduces weighting and validation by the scientific community into the corpus of our surveyed literature.

We are aware that the selection of keywords has an impact on the selection of potential publications.
However, we argue that the common vocabulary of counterexample explanation is represented in our list of keywords. The usage of more general keywords such as ``formal methods'' would increase the initial count of publications at the cost of drastically increasing false positives beyond a reasonable amount. Therefore, we combined general keywords with the keywords that are specific to counterexample explanation.

As another means to increase external validity, we performed snowballing with the publications that we filtered from the initial search. This process potentially incorporates literature that is relevant to the field but is not matched by our keyword list.

In summary, we are confident that the selected literature for this survey is representative for its community, which minimizes threats to external validity.

\subsection{Reliability}
\label{sec:ttv:reliability}

\noindent
To ensure reliability of the analysis results presented in \sect{analysis}, we applied a four-eyes principle. The literature obtained by our selection process was collected in a table, shared and worked on by all authors of this survey.

For each of the research questions documented in \sect{RQ}, relevant data for the quantitative RQs was collected, reviewed, and aggregated. We applied descriptive statistics on the aggregated data to investigate the quantitative aspects of the research questions.
For each of the qualitative aspects, relevant statements were collected in the shared table and reviewed as well. These statements were clustered, summarized, and exemplary statements were chosen for discussion.

We are confident that this approach minimizes threats to the reliability of the presented results.

%% file: tex/Discussion.tex
\section{Discussion}
\label{sec:discussion}

\noindent
Based on the survey of the literature, we gained significant knowledge and understanding in the domain of counterexample explanation. In the following, we use this knowledge and understanding to discuss the main findings of our study, and suggest future research directions for the community.

\subsection{Synopsis}
\label{sec:discussion:synopsis}

\noindent
In this survey, we address key concerns in model checking for error comprehension along the research questions introduced in \sect{RQ}. The result of the analysis shall serve as a reference for the research community, giving insights to existing research approaches. We surveyed publications since the early 2000's, indicating that counterexample interpretation is an active research field for at least two decades by now.

\subsubsection{Counterexample explanations for different audiences and usage scenarios}

\noindent
We discuss counterexample explanations from the perspective of three different users:
\begin{enumerate*}[label={(\textbf{U\arabic*})},itemjoin={{, }}, itemjoin*={{, and }}]
	\item experts with strong knowledge of both the formal methods and the domain
	\item domain experts with limited formal methods knowledge
	\item domain experts without formal methods knowledge.
\end{enumerate*}

Research question \rsqtext{1} (\sect{analysis:representation}) mainly focuses on four different types of counterexample representation: textual, graphical, trace, and tabular notation. These different representations are tailored to users based on their experience, knowledge, and background.
Counterexample representation as a trace or tabular notation seems to be most suitable for user \textbf{U1} as it is still close to the raw output of the model checkers. Tabular notations convert a counterexample into rows and columns. To obtain a trace representation, the counterexample is processed and provided as a trace.

Some of the graphical counterexample representation formats discussed in the context of \rsqtext{4} (\sect{analysis:domain}), \eg fault trees, function block diagrams, and component diagrams, as well as representations inside a programming language seem to be suited for user \textbf{U2}. For instance, compared to a lengthy counterexample, compact and concise fault tree representation is efficient and accustomed to safety engineers, easing comprehension of errors. Similarly, function block diagrams and component diagram are suitable for system engineers or system developers just as programming languages are for software engineers and programmers.

Additionally, with \rsqtext{4} (\sect{analysis:domain}) we found that a representation of a counterexample within the input domain is easy for a domain experts to comprehend, \eg by reference/mapping and particularly by simulation. The same argument is made by Loer and Harrison~\cite{LoerH06}, who argue that \enquote{to be useful, the results of the analysis need to be visualized in a way that the designer can use. Traces can form the backbone for scenarios that may be analyzed by requirements engineers. They can be used to illustrate and communicate the implications of design decisions to and between design teams.}

In most of the primary studies, counterexamples are represented in a graphical format. While such graphical visualization is useful mainly for user \textbf{U1} and \textbf{U2}, it may not provide enough support for user \textbf{U3}. To overcome this, textual explanations of counterexamples seem to be a promising direction, \eg expressed in a \gls{CNL} or structured English. These explanations may be enriched by domain ontologies, domain vocabularies, and domain-specific terminology to improve their value. This type of language does not require any formal methods knowledge, thus it is suitable for user \textbf{U3}. 
Furthermore and as discussed with \rsqtext{2} (\sect{analysis:processed}) and \rsqtext{3} (\sect{analysis:enrich}), natural language like textual representation along with additional information like error highlighting or error localization in the given input source, improves error comprehension and supports user \textbf{U3} in analysis and debugging.

\subsubsection{Counterexample processing}

\noindent
Traces are one form of representing counterexamples, which we discussed with \rsqtext{2} (\sect{analysis:processed}) and \rsqtext{3} (\sect{analysis:enrich}). It is further categorized into witness traces, minimized counterexamples, and multiple counterexamples. According to Aljazzar \etal~\cite{AljazzarLLS11}, the shortest diagnostic paths, \ie the paths with least transitions, are preferred over longer paths. Copty \etal~\cite{CoptyIWKK03} state that using multiple counterexamples allows to find all possible failures in a single verification run, and has the ability to identify more than one root cause so that it can reduce the number of verification runs.

From our point of view, multiple counterexamples can be useful for experts to understand an error in a detailed and time-efficient manner. However, as per the statement from Aljazzar \etal~\cite{AljazzarLLS11}, reduction of counterexamples by removing unwanted paths, states, and variables can make the error comprehension easier when compared to concrete counterexamples.

Having witness traces along with a counterexample is an added advantage, which we discussed with \rsqtext{2} (\sect{analysis:processed}) and \rsqtext{3} (\sect{analysis:enrich}) in detail. Witness traces are proofs that highlights the behavior of the model. It is used to justify the result of the model checker, which are contrary to the counterexample. Comparing witness traces with counterexamples can aid in identification of the erroneous behavior in a counterexample~\cite{ShenQL04,KumazawaT11}.

Moreover, with \rsqtext{2} (\sect{analysis:processed}) we summarized different methods, techniques, or approaches that are used to process counterexamples. Most of these approaches are used for counterexample minimization. In our view, integration of model transformation with model checker promises to be an efficient method because it performs both model abstraction and counterexample representation when transforming design models to verification models and vice versa.
For instance, works by  Ratiu \etal~\cite{RatiuGS19}, Campetelli \etal~\cite{CampetelliJBDKZ15}, Gerking \etal~\cite{GerkingSDH15}, and Luteberget \etal~\cite{LutebergetCJS17} present comprehensive frameworks that perform both the model transformation and verification. Such frameworks hide the formal models and visualize relevant information from the counterexample in the user-provided design model of the system.

\subsubsection{Frameworks and tools}

\noindent
Looking at the quantitative results of \rsqtext{6} (\sect{analysis:tools}), frameworks for counterexample representation do not seem to be reused in a regular manner. Generally, if a framework is found in multiple publications, these are usually quite homogeneous research groups around the framework's authors. Only a few publications were found that have re-used existing third-party frameworks.

The same is true for verification tools. Even though the large number of verification tools we found in the literature, only a few verification tools such as \gls{NuSMV}, \gls{SPIN}, Maude, and \gls{PRISM} have a wide adoption. \gls{NuSMV} is the most used verification tool in the surveyed literature. A reason might be that NuSMV supports different types of model checking, model encoding techniques, and input domains.

\subsubsection{Application and evaluation}

\noindent
With \rsq{7} (\sect{analysis:evaluation}), we discussed evaluation aspects, methods, types of applications, and application domains. We would like to highlight that the surveyed approaches were evaluated with a considerable number of industrial applications in safety-critical domains such as automotive, nuclear industries, and avionics.
We show that the preferred counterexample representation depends on the application domain. For certain domains, graphical representation is more feasible, while for other domains textual representation is best suited. For example, the textual representation from Berg \etal~\cite{BergSJ07} and Luteberget \etal~\cite{LutebergetCJS17,LutebergetJ18} is found to be efficient for the railway domain, while fault tree representation is found to be most suited for safety engineers, \eg in the automotive domain~\cite{KuntzLL11}.

\subsection{Future directions}
\label{sec:analysis:future}

\noindent
This survey reveals several open points and future directions to enhance counterexample explanation approaches. Referring to the quantitative results of \rsqtext{1}, there is a considerable amount of work that uses graphical representations of counterexamples. Still, approaches focusing on textual representations as well as representations suited for laypersons (\textbf{U3}) are to be improved. \change{Similarly, the quantitative results of \rsqtext{2} show that a considerable amount of work addresses counterexample minimization and its visualization. However, work focusing on multiple counterexamples particularly, effective visualization of multiple counterexamples and highlighting relevant sources for debugging of multiple counterexamples seems to provide potential for further research. Such research could support comprehending multiple counterexamples for experts in formal methods (\textbf{U1}) and domain experts with limited knowledge in formal methods (\textbf{U2}).}

Looking at the quantitative results for verification tools and temporal logics in \rsqtext{5} and \rsqtext{6}, qualitative model checkers such as \gls{NuSMV} and \gls{SPIN}, and qualitative logics such as \gls{LTL} and \gls{CTL} are used predominantly. Although there are several challenges in using probabilistic model checkers as discussed with \rsqtext{6}, in this survey we found a considerable number of publications using probabilistic model checkers. However, the number of approaches using real-time model checkers is very low, which indicates room for further research. Similarly, referring to \rsqtext{5} for specification properties, there are plenty of contributions that address safety properties, leaving room for research on counterexample explanation when verifying liveness properties, especially considering the increase in autonomy of modern systems.

We found many frameworks for counterexample explanation (\rsqtext{6}) and a considerable number of these are open-source. Therefore, it seems advisable to build upon existing frameworks and improve them. Also, a large number of publications address either processed counterexamples \change{or} counterexample representation. In our view, developing an integrated framework that could perform both: the processing of counterexample and representation of it might be a powerful tool that would help practitioners in adopting and using counterexample explanation approaches.

To improve the effectiveness and usability of counterexample explanations, user studies need to be performed to gain insights in the degree of error comprehension and understandability, the key concerns of counterexample explanation. In our survey, though, we only found a small number of user studies (see \rsqtext{7}). Therefore, researchers should consider performing user studies that allow investigating the error comprehension and understanding that users obtain when using counterexample explanation approaches. Such user studies will provide evidence for the effectiveness and usability of such approaches from a user's point of view and thus can guide future research on counterexample explanation.

%% file: tex/conclusion.tex
\section{Conclusion}
\label{sec:conclusion}

\begin{center}
	\blockquote{\textit{As researchers and educators in formal methods, we should strive to make our notations and tools accessible to non-experts.} -- Edmund Clarke~\cite{ClarkeW96}}
\end{center}

\noindent
This article surveys the available literature on counterexample explanation, which addresses key concerns in model checking for error comprehension, along with a set of quantitative and qualitative research questions. Analyzing \pubfinal primary studies we identified in this survey, the quantitative results are provided for different types of counterexample representations, counterexample processing, model checkers, temporal logics, frameworks, applications, application domains, evaluation aspects, and evaluation methods.
Further, from a qualitative viewpoint, we discuss aspects such as methods and tools used in counterexample explanation approach as well as their effects, dependencies, and challenges for the explanations. As a result, this systematic literature review shall serve as a reference for the research community, giving insights to existing research approaches focusing on counterexample explanation. To the best of our knowledge, our systematic literature review is the first to provide such a reference for counterexample explanation.